%% file: paper.tex
\documentclass[a4paper,twoside,12pt]{article}

\usepackage{a4wide}
\usepackage[tbtags]{amsmath}
\usepackage{amssymb}
\usepackage{amsthm}
\usepackage{fancyhdr}
\usepackage[dvips]{graphicx}
\usepackage{exscale}
\usepackage{ifthen}
\usepackage{color}
\usepackage[bookmarks]{hyperref}

\newcommand{\sh}[1]{
\ifthenelse{\equal{#1}{l}}
  {
   \text{$\not{#1}$}
  }
  {
   \text{$\not{\!#1}$}
  }
}
\newcommand{\lqcd}{\Lambda_{\text{QCD}}}
\newcommand{\Op}{\mathcal{O}}
\newcommand{\intd}{\int\frac{d^dk}{(2\pi)^d}}

\newcommand{\lmu}{\ln\frac{\mu^2}{m_b^2}}
\newcommand{\li}{\text{Li}_2}
\theoremstyle{definition}
\newtheorem{ex}{Example}[section]

\allowdisplaybreaks[1]

\title{Hard spectator interactions in $B\to\pi\pi$ at order $\alpha_s^2$}

\date{}

\begin{document}
\hfill LMU-ASC 68/07\\
\begin{center}
{\LARGE Hard spectator interactions in $B\to\pi\pi$ at order
  $\alpha_s^2$}
\\[4ex]
{\small Volker Pilipp\footnote{volker.pilipp@itp.unibe.ch}\\
\footnotesize Arnold Sommerfeld Center, Department f\"ur Physik\\
\footnotesize Ludwig-Maximilians-Universit\"at M\"unchen\\
\footnotesize Theresienstrasse 37, 80333 M\"unchen, Germany\\[3ex]
\footnotesize Institute of Theoretical Physics\\
\footnotesize Universit\"at Bern\\
\footnotesize Sidlerstrasse 5, 3012 Bern, Switzerland
}
\end{center}
\begin{abstract}
I compute the \emph{hard spectator interaction}
amplitude in $B\to\pi\pi$ at NLO i.e.\ at $\mathcal{O}(\alpha_s^2)$. 
This special part of
the amplitude, whose LO starts at $\mathcal{O}(\alpha_s)$, is defined in
the framework of QCD factorization. QCD factorization allows to separate the 
short- and the long-distance physics in leading power in an expansion
in $\lqcd/m_b$, where the short-distance physics can be
calculated in a perturbative expansion in $\alpha_s$.

In this calculation it is necessary to
obtain an expansion of Feynman integrals in powers of
$\Lambda_\text{QCD}/m_b$. I will present a general method to obtain this
expansion in a systematic way once the leading power is given as an
input. This method is based on differential equation techniques and
easy to implement in a computer algebra system.

The numerical impact on amplitudes and branching ratios is
considered. The NLO contributions of the hard spectator interactions 
are important but small enough for perturbation theory to be valid.
\end{abstract}
\section{Introduction}
\input{qcdfact/qcdfact}
\section{Notation and basic formulas\label{notbf}}
\input{basic/basic}

\section{Hard spectator interactions at LO \label{HspLO}}
\input{LO/LO}

\section{Calculation of Feynman diagrams with differential 
equations \label{diffeq}}
\input{diffeq/diffeq}
\section{Technical details of the NLO calculation\label{tech}}
\input{diagrams/diagrams_farxiv}

\input{wavefunctions/wf}

\section{NLO results \label{nloresults}}
\subsection{Analytical results for $T^\text{II}_1$ and  $T^\text{II}_2$}
 \input{result/result}

\subsection{Convolution integrals and factorizability}
\input{result/convint}
\section{Numerical analysis\label{numres}}
\input{result/num}

\section{Conclusions}
\input{conclusions/conclusions}

\section*{Acknowledgements}

I would like to thank Gerhard Buchalla for 
proofreading the drafts and comments on the manuscript.
I am grateful to Guido Bell, Matth\"aus Bartsch and 
Sebastian J\"ager for many instructive and helpful discussions.

\begin{appendix}
 \section{Massive four-point integral}
 \label{4pointmass}
 \input{fourpoint/fourpoint}
\end{appendix}

\input{bib}
\end{document}

%% file: qcdfact/qcdfact.tex
In the last decades $B$ physics has proven to be a promising field to
determine parameters of the flavour sector with high precision. On the
theoretical side QCD factorization \cite{Beneke:1999br,Beneke:2000ry}
has turned out to be an appropriate tool to calculate $B$ decay modes
from first principles. 
Though the decay of the $B$-meson is caused by weak interactions,
strong interactions play a dominant role. It is however not possible
to handle the QCD effects completely perturbatively. This is due to
the energy scales that are contained in the $B$-meson: Whereas $\alpha_s$ 
at the mass of the $b$-quark is a small parameter, the bound
state of the quarks leads to an energy scale of $\mathcal{O}(\lqcd)$
which spoils perturbation theory. The idea of QCD factorization
is to separate these scales. At
leading power in $\lqcd/m_b$ we obtain the amplitude for $B\to\pi\pi$ in
the following form:
\begin{eqnarray}
  \langle \pi\pi| \mathcal{H} | B \rangle & \sim & 
  F^{B\to\pi}\int_0^1 dx\,T^\text{I}(x)f_\pi\phi_\pi(x)+ \nonumber \\
  &&\int_0^1 dx dy d\xi \, T^\text{II}(x,y,\xi) 
    f_B\phi_{B1}(\xi)f_\pi\phi_\pi(x)f_\pi\phi_\pi(y)
   \label{factform}
\end{eqnarray}
Two different types of quantities enter this formula. On the one hand
the hadronic physics is contained in the form
factor $F^{B\to\pi}$ and the wave functions $\phi_{B1}$ and $\phi_\pi$,
which will be defined more precisely in the next section. These
quantities contain the information about the bound states of the
mesons. They have to be determined by non-perturbative methods like QCD 
sum rules or lattice calculations. Alternatively, because they are at least
partly process independent, they might be extracted in the future from
experiment. On the other hand the hard scattering kernels $T^\text{I}$
and $T^\text{II}$ contain the physics of the hard scale $\mathcal{O}(m_b)$ and 
the hard collinear scale $\mathcal{O}(\sqrt{m_b\lqcd})$ and can be
calculated perturbatively. 

\begin{figure}[t]
\begin{center}
\resizebox{\textwidth}{!}{\includegraphics{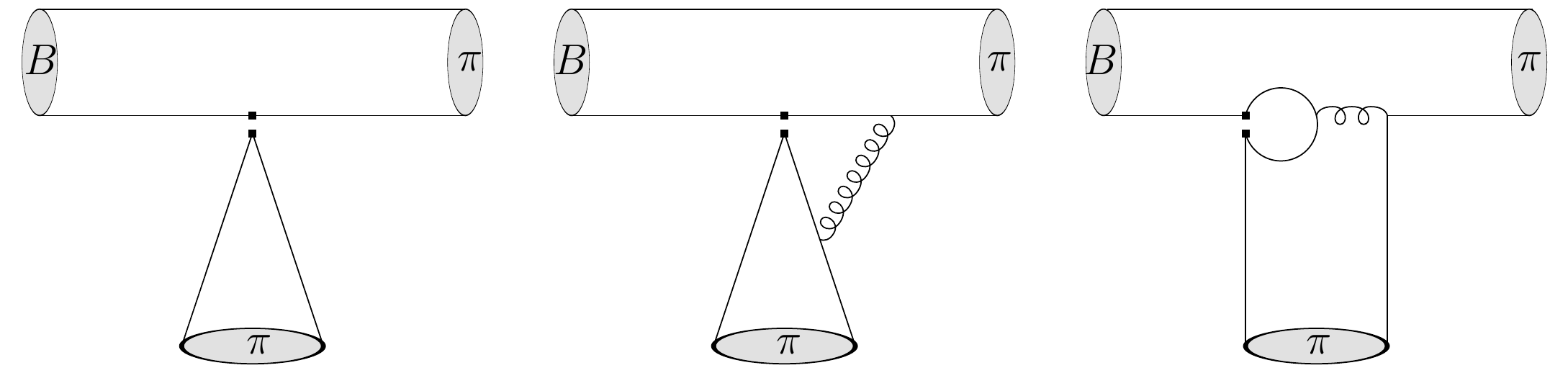}}
\end{center}
\caption{Tree level, vertex correction and penguin contraction. 
         These diagrams contribute to $T^\text{I}$.}
\label{vertex}
\end{figure}

\begin{figure}[t]
\begin{center}
\resizebox{\textwidth}{!}{\includegraphics{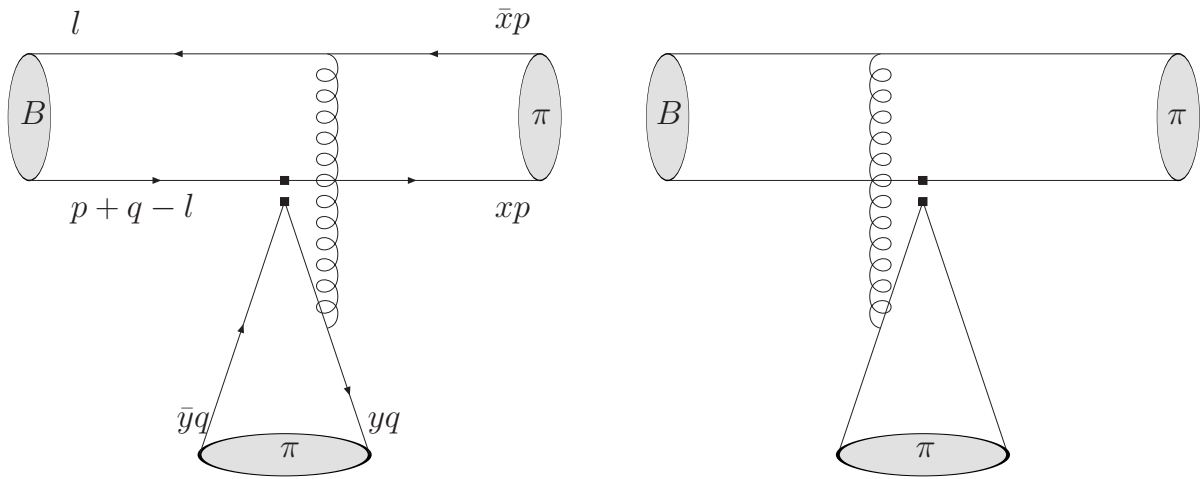}}
\end{center}
\caption{Hard spectator interactions at $\Op(\alpha_s)$. This is the
  LO of $T^\text{II}$}
\label{basic}
\end{figure}

The Feynman diagrams that
contribute to $B\to\pi\pi$ can be distributed into two different
classes. 
The class of diagrams where there is no gluon line connecting the
spectator quark with the rest of the diagram (fig.~\ref{vertex}) 
contributes to $T^\text{I}$.
We obtain $T^\text{II}$ by evaluating the hard spectator scattering
diagrams, which are shown in LO in $\alpha_s$ in fig.~\ref{basic}.
The order $\alpha_s^2$ corrections of $T^\text{II}$ are the topic of
the present work.
Through the soft momentum $l$ of the constituent quark of the
$B$-meson the hard collinear
scale $\sqrt{\lqcd m_b}$ comes into play. This leads to the fact 
that in contrast to $T^\text{I}$, which is completely governed by the 
scale $m_b$, $T^\text{II}$ comes with formally large logarithms. These
logarithms cannot be resummed in the present QCD calculation. It will
be shown by numerical analysis that the scale dependence of the hard
spectator scattering amplitude and the absolute size of its NLO
corrections are small enough for pertubation theory to be
valid.


My calculation of the hard spectator scattering
amplitude is not the first one as it has been calculated recently
by \cite{Beneke:2005vv,Kivel:2006xc}. It is however the first
pure QCD calculation, whereas \cite{Beneke:2005vv,Kivel:2006xc}
used the framework of soft-collinear effective theory (SCET)
\cite{Bauer:2000yr,Bauer:2001yt,Beneke:2002ph} an effective theory, 
where the expansion in $\lqcd/m_b$ is performed at the level of the 
Lagrangian rather than of Feynman integrals. It is the main result 
of this paper to confirm the results of
\cite{Beneke:2005vv,Kivel:2006xc} and to show by explicit calculation 
that pure QCD and SCET lead to the same result in this case. 

Hard spectator
scattering corrections to the penguin diagram (third diagram of
fig.~\ref{vertex}) are beyond the scope of this publication. This
class of diagrams does not influence the cancellation of the scale
dependence of the 
``tree amplitude'' i.e.\ the diagrams of fig.~\ref{basic} and 
higher order $\alpha_s$ corrections. For phenomenological applications, 
however, the hard spectator penguin amplitudes, which have been
recently calculated in \cite{Beneke:2006mk}, should be taken into 
account. 
Also the order $\alpha_s^2$ of $T^\text{I}$ is important for
phenomenological applications. This calculation has been partly performed by
\cite{Bell:2007tz,Bell:2007tv}. There the complete imaginary part
and a preliminary result of the real part of the amplitude is given.

From a technical point of view the calculation in this paper consists
of the evaluation of about 60 one-loop Feynman diagrams. The
challenges of this task are due to the fact that these diagrams come
with up to five external legs and three independent ratios of
scales. In order to reduce the number of master integrals and to
perform power expansions of the Feynman integrals, integration by parts
methods and differential equation techniques will prove appropriate
tools. They provide a general method to obtain higher powers of a Feynman
integral once the leading power is given.

The paper is organized as follows: I define my notations in section
\ref{notbf}. In section \ref{HspLO} I show how to get $T^\text{II}$ at
LO. 
In section \ref{diffeq} I present a method which uses
differential equation techniques and allows for the extraction of higher
powers of Feynman integrals once the leading power is given.
Sections \ref{tech} is dedicated to the technical details of the
calculation.  After
some remarks how to evaluate the Feynman diagrams occuring at NLO I
show how to deal with meson wave functions at NLO. Especially the
correct treatment of evancescent structures will be explained in
detail. In section \ref{nloresults} the analytic results of the
calculation will be given, whereas section \ref{numres} will provide
the numerical analysis. I end up with the conclusions.

%% file: basic/basic.tex
\subsection{Kinematics}
For the process $B\to\pi\pi$ we will assign the momenta $p$ and
$q$ to the pions (fig.~\ref{basic}) which fulfil
the condition 
\begin{equation}
p^2,q^2=0.
\label{kin1}
\end{equation}
This is the leading power approximation in $\Lambda_\text{QCD}/m_b$. 
Let us define two Lorentz vectors $n_+$, $n_-$ by:
\begin{equation}
n_+^\mu\equiv(1,0,0,1),\quad n_-^\mu\equiv(1,0,0,-1).
\label{kin2}
\end{equation}
In the rest frame of the decaying meson $p$ can be defined to be in the
direction of $n_+$ and $q$ to be in the direction of $n_-$. 
Light cone coordinates for the Lorentz vector $z^\mu$ are defined by:
\begin{equation}
z^+\equiv\frac{z^0+z^3}{\sqrt{2}},\quad
z^-\equiv\frac{z^0-z^3}{\sqrt{2}},\quad
z_\perp\equiv (0,z^1,z^2,0)
\label{k5}\end{equation}
So one can decompose $z^\mu$ into:
\begin{equation}
z^\mu=\frac{z\cdot p}{p\cdot q} q^\mu+\frac{z\cdot q}{p\cdot q}
p^\mu+z_\perp^\mu \label{kin3}
\end{equation}
such that 
\begin{equation}
z_\perp\cdot p=z_\perp\cdot q =0.\label{kin4}
\end{equation}

\subsection{Colour factors}
In our calculations we will use the following three colour factors,
which arise from the $\text{SU}(3)$ algebra:
\begin{equation}
  C_N=\frac{1}{2},\quad C_F=\frac{N_c^2-1}{2N_c}\quad\text{and}\quad
  C_G=N_c,
  \label{colorfactor}
\end{equation}
where $N_c=3$ is the number of colours. 

\subsection{Meson wave functions}
The pion light cone distribution amplitude $\phi_\pi$ is defined by
\begin{equation}
  \label{mw1}
  \langle\pi(p)|\bar{q}(z)_\alpha [\ldots] q^\prime(0)_\beta|0\rangle_{z^2=0} =
  \frac{if_\pi}{4}(\sh{p}\gamma_5)_{\beta\alpha}
  \int_0^1 dx\, e^{ix p\cdot z}\phi_\pi(x).
\end{equation}
The ellipsis $[\ldots]$ stands for the Wilson line 
\begin{equation}
[z,0]=\text{P}\exp\left(\int_0^1 dt\,ig_s z\cdot A(z t)\right),
\end{equation}
which makes (\ref{mw1}) gauge invariant.
For the definition of the $B$-meson wave function $\phi_{B1}$ we need
the special kinematics of the process. Following \cite{Beneke:2000ry} let us 
define
\begin{equation}
\Psi_B^{\alpha\beta}(z,p_B)=
\langle 0| \bar{q}_\beta(z)[\ldots]b_\alpha(0)|B(p_B)\rangle
=\int\frac{d^4l}{(2\pi)^4}e^{-il\cdot z}\phi_B^{\alpha\beta}(l,p_B).
\label{mw2}\end{equation}
In the calculation of matrix elements we get terms like:
\begin{equation}
\int\frac{d^4l}{(2\pi)^4}\text{tr}(\mathcal{A}(l)\phi_B(l))=
\int\frac{d^4l}{(2\pi)^4}\int d^4z\, e^{il\cdot z}\text{tr}
(\mathcal{A}(l)\Psi_B(z)).
\label{mw3}
\end{equation}
We have to consider only the case that the amplitude $\mathcal{A}$
depends on $l$ only through $l\cdot p$:
\begin{equation}
\mathcal{A}=\mathcal{A}(l\cdot p)
\end{equation}
In this case we can use the $B$-meson wave function on the light cone
which is given by \cite{Beneke:2000ry}:
\begin{eqnarray}
\lefteqn{\langle 0|
\bar{q}_\alpha(z)[\ldots]b_\beta(0)|B(p_B)\rangle\bigg|_{z^-,z_\perp=0}}
\label{mw4}\\
&&=-\frac{if_B}{4}[(\sh{p}_B+m_b)\gamma_5]_{\beta\gamma}\int_0^1 d\xi\,
e^{-i\xi p_B^-
z^+}[\Phi_{B1}(\xi)+\sh{n}_+\Phi_{B2}(\xi)]_{\gamma\alpha}\nonumber
\end{eqnarray}
where
\begin{equation}
\int_0^1 d\xi\,\Phi_{B1}(\xi)=1\quad\mbox{and}\quad 
\int_0^1 d\xi\,\Phi_{B2}(\xi)=0.
\label{mw5}\end{equation}
It is now straightforward to write down the momentum projector of the
$B$-meson:
\begin{eqnarray}
\lefteqn{
\int\frac{d^4 l}{(2 \pi)^4}\, \text{tr} (A(2 l\cdot p)\hat{\Psi}(l))}
\nonumber\\
&&=\frac{-i f_B}{4}\text{tr}(\sh{p_B}+m_B)\gamma_5
\int_0^1 d\xi\,(\Phi_{B1}(\xi)+\sh{n}_+\Phi_{B2}(\xi))A(\xi m_B^2)
\label{mw6}
\end{eqnarray}
At this point we give the following definitions
\begin{eqnarray}
\frac{m_B}{\lambda_B}&\equiv&\int_0^1\frac{d\xi}{\xi}\phi_{B1}(\xi)
\label{lambdaB}\\
\lambda_n &\equiv& \frac{\lambda_B}{m_B}
\int_0^1\frac{d\xi}{\xi}\ln^n\xi\phi_{B1}(\xi).
\label{lambdan}
\end{eqnarray}

%% file: LO/LO.tex
The effective weak Hamiltonian we deal with is given by 
\cite{Buchalla:1995vs}:
\begin {equation}
\mathcal{H}_\text{eff}=\frac{G_F}{\sqrt{2}}
V_{ud}^*V_{ub}
\left[C_1 \mathcal{O}_1 + C_2 \mathcal{O}_2
\right]+\text{h.c.}, 
\label{w0}\end{equation}
where 
\begin{eqnarray}
\Op_1 & = & (\bar{d}p)_{V-A}(\bar{p}b)_{V-A},\nonumber\\
\Op_2 & = & (\bar{d}_ip_j)_{V-A}(\bar{p}_jb_i)_{V-A}.\label{w8}
\end{eqnarray}
Explicit expressions for the short-distance coefficients $C_i$ can be
obtained from \cite{Buchalla:1995vs}. 
The decay amplitude of $B\to\pi\pi$ is given by
\begin{equation}
\mathcal{A}(B\to\pi\pi)\equiv\langle\pi\pi|\mathcal{H}_\text{eff}|B\rangle.
\label{w9}
\end{equation}
For later convenience we define 
\begin{equation}
\mathcal{A}(B\to\pi\pi)\equiv
\mathcal{A}(B\to\pi\pi)^\text{I}+\mathcal{A}(B\to\pi\pi)^\text{II}
\label{w10}
\end{equation}
where 
$\mathcal{A}^\text{I}$ ($\mathcal{A}^\text{II}$) belongs to the first (second)
term of (\ref{factform}). Because $\mathcal{A}^\text{I}$ and
$\mathcal{A}^\text{II}$ contain different hadronic quantities, the
renormalisation scale dependence of both of them has to vanish separately. 
So we can set their scales to different values $\mu^\text{I}$ and
$\mu^\text{II}$. As in $\mathcal{A}^\text{I}$ there occurs only the 
mass scale $m_b$
we can set $\mu^\text{I}=m_b$. In $\mathcal{A}^\text{II}$ there occurs also the
hard-collinear scale $\sqrt{\lqcd m_b}$.  As we will see this scale is an
appropriate choice for $\mu^\text{II}$.

Because we only deal with the tree amplitude and do not consider
penguin contractions only the matrix elements of the 
operators $\Op_1$ and $\Op_2$ are taken into account. These operators
close under renormalisation such that the corresponding hard spectator 
amplitude is independent of the renormalisation scale.

The decay amplitudes of
$B\to\pi\pi$ can be written in terms of $a_i$ as follows \cite{Beneke:2001ev}:
\begin{eqnarray}
-\mathcal{A}(\bar{B}^0\to\pi^+\pi^-) &=&
\left[\lambda_u^\prime a_1+\lambda_p^\prime(a_4^p+r_\chi^\pi
  a_6^p)\right]A_{\pi\pi}
\nonumber\\
-\sqrt{2}\mathcal{A}(B^-\to\pi^-\pi^0) &=&
\lambda_u^\prime(a_1+a_2)A_{\pi\pi}
\nonumber\\
\mathcal{A}(\bar{B}^0\to\pi^0\pi^0) &=& 
\left[
-\lambda_u^\prime a_2+\lambda_p^\prime(a_4^p+r_\chi^\pi a_6^p)
\right]A_{\pi\pi}
\label{w13}
\end{eqnarray}
where 
$$A_{\pi\pi}=i\frac{G_F}{\sqrt{2}}(m_B^2-m_\pi^2)f^{B\pi}_+f_\pi$$
and 
\begin{equation}
r_\chi^\pi(\mu)=\frac{2m_\pi^2}{\bar{m}_b(\mu)(\bar{m}_u(\mu)+\bar{m}_d(\mu))}.
\label{w14}
\end{equation}
For the LO and NLO results of the $a_i$ I refer to
\cite{Beneke:2001ev}. We define analogously to (\ref{w10})
\begin{equation}
a_i=a_{i,\text{I}}+a_{i,\text{II}}
\label{w15}
\end{equation}
where the labels `I' and `II' refer to the contribution to
$\mathcal{A}^\text{I}$ and $\mathcal{A}^\text{II}$.

The leading order of the hard spectator interactions which start at
$\mathcal{O}(\alpha_s)$  is shown in \mbox{fig.~\ref{basic}}. 
The hard spectator scattering kernel
$T^\text{II}$, which does not depend on the wave functions, can be
obtained by calculating the transition matrix element between free
external quarks, to which we assign the momenta shown in
\mbox{fig.~\ref{basic}}. The variables $x, \bar{x}\equiv 1-x, y,
\bar{y}\equiv 1-y$ are the arguments of $T^\text{II}$, which arise from the
projection on the pion wave function (\ref{mw1}). In the sense of
power counting we count all components of $l$ of 
$\mathcal{O}(\lqcd)$, while the components of $p$ and $q$ are 
$\mathcal{O}(m_b)$ or
exactly zero. We define the following quantities
\begin{equation}
\xi      \equiv  \frac{l\cdot p}{p\cdot q}, \quad\quad 
\theta   \equiv  \frac{l\cdot q}{p\cdot q}. \label{LO1}
\end{equation} 
We will see that in the end the dependence on $\theta$ vanishes 
in leading power such that we can use (\ref{mw6}).

We consider the three cases $\bar{B}^0\to\pi^+\pi^-$,
$\bar{B}^0\to\pi^0\pi^0$ and  $B^-\to\pi^-\pi^0$. 
In the case, that the external quarks come with the flavour content of 
$\bar{B}^0\to\pi^+\pi^-$, the LO hard spectator 
amplitude for the effective operator $\Op_2$ reads:
\begin{eqnarray}
\lefteqn{A_\text{spect.}^{(1)}(\bar{B}^0\to\pi^+\pi^-)\equiv}\nonumber\\
&&\langle \bar{d}(\bar{x}p) u(xp)\, \bar{u}(\bar{y}q)d(yq)|
\Op_2 | \bar{d}(l)b(p+q-l)\rangle_\text{spect.}=\nonumber\\
&&4\pi\alpha_s C_F N_c\frac{1}{\bar{x}\xi m_B^2}
\bar{d}(l)\gamma^\mu d(\bar{x} p)\,
\bar{u}(x p) \gamma^\nu(1-\gamma_5)b(p+q-l)\nonumber\\
&&\bar{d}(y q)\left(
\frac{2\sh{p}g_{\mu\nu}}{\bar{y}}-
\frac{\sh{p}}{y\bar{y}}\gamma_\mu\gamma_\nu\right)(1-\gamma_5)
u(\bar{y}q),
\label{LO2}
\end{eqnarray}
where the quark antiquark states in the input and output channels of
the matrix element form colour singlets. The subscript ``spect.''\ 
means that only diagrams with a hard spectator interaction are taken
into account. The amplitude of $\Op_1$ vanishes to this order in 
$\alpha_s$. In the case of 
$\bar{B}^0\to\pi^0\pi^0$ we get the tree amplitude from the matrix
element of $\Op_1$. The case $B^-\to\pi^-\pi^0$ does not need to
be considered separately, because from isospin symmetry follows
\cite{Gronau:1994rj,Beneke:2001ev}:
\begin{equation}
\sqrt{2}\mathcal{A}(B^-\to\pi^-\pi^0) =
\mathcal{A}(\bar{B}^0\to\pi^+\pi^-)+
\mathcal{A}(\bar{B}^0\to\pi^0\pi^0).
\end{equation}

On the other hand the full amplitude is the convolution of
$T^\text{II}$ with the wave functions, given by (\ref{factform}). 
To extract $T^\text{II}$ from
(\ref{LO2}) we need the wave functions with the same external states we
have used in (\ref{LO2}),
i.e.\ we have to calculate the matrix elements (\ref{mw1}) and
(\ref{mw2}), where the pion or $B$-meson states are replaced by free
external quark states. To the order $\mathcal{O}(\alpha_s^0)$ we get
\begin{eqnarray}
\phi_{\pi^-\alpha\beta}^{(0)}(y^\prime) &\equiv&
\int d(z\cdot q) e^{-i z\cdot q y^\prime}
\langle \bar{u}(\bar{y} q) d(y q)|\bar{d}^i_\beta(z) u^i_{\alpha}(0)|
0\rangle_{z^-,z_\perp=0} \nonumber\\
&=& 2\pi N_c\delta(y^\prime-y)
\bar{d}_\beta(yq)u_\alpha(\bar{y}q)\nonumber\\
\phi_{\pi^+\alpha\beta}^{(0)}(x^\prime) & = &
2\pi N_c\delta(x^\prime-x)\bar{u}_\beta(xp)d_\alpha(\bar{x}p)\label{LO3}\\
\phi_{B\alpha\beta}^{(0)}(l^{\prime -}) &\equiv& 
\int dz^+ e^{il^{\prime -} z^+}
\langle 0|\bar{d}_\beta(z)^i b_\alpha(0)^i|\bar{d}(l)b(p+q-l)
\rangle_{z^-,z_\perp=0}\nonumber\\
&=&2\pi N_c\delta(l^{\prime -}-l^-)\bar{d}_\beta(l)b_\alpha(p+q-l)
\nonumber
\end{eqnarray}
By using 
\begin{equation}
A^{(1)}_\text{spect.}=\int dxdydl^-\, 
\phi^{(0)}_{\pi^+\alpha\alpha^\prime}(x)
\phi^{(0)}_{\pi^-\beta\beta^\prime}(y)
\phi^{(0)}_{B\gamma\gamma^\prime}(l^-)
T^{\text{II}(1)}(x,y,l^-)_{\alpha^\prime\alpha\beta^\prime\beta\gamma^\prime\gamma}
\label{LO3.1}
\end{equation}
we finally obtain:
\begin{eqnarray}
T^{\text{II}(1)}(x,y,l^-)_{\alpha^\prime\alpha\beta^\prime\beta\gamma^\prime\gamma}
&=&
4\pi\alpha_s \frac{C_F}{(2\pi)^3 N_c^2}\frac{1}{\xi\bar{x}m_B^2}
\gamma^\mu_{\gamma^\prime\alpha}
\left[\gamma^\nu(1-\gamma_5)\right]_{\alpha^\prime\gamma}
\nonumber\\
&&\left[\left(\frac{2\sh{p}g_{\mu\nu}}{\bar{y}}-
\frac{\sh{p}}{y\bar{y}}\gamma_\mu\gamma_\nu\right)(1-\gamma_5)\right]_
{\beta^\prime\beta}.
\label{LOkernel}
\end{eqnarray}
It should be noted that only the first summand of the above equation 
contributes after performing the Dirac trace in four dimensions. The
second summand is evanescent. This will be important, when we will
calculate the NLO corrections of the wave functions (see section
\ref{wf}).
 
If we plug the hadronic wave functions defined by (\ref{mw1}) 
and (\ref{mw6}) into (\ref{LO3.1}) i.e.\ we calculate the matrix element 
(\ref{LO2}) between meson states instead of free quark states,
we get for the LO amplitude
\footnote{$A_\text{spect.}$ is used for the
  matrix elements of the operators $\Op_i$ between both free
  external quarks and hadronic meson states. It should become clear
  from the context what is actually meant.}:
\begin{equation}
A^{(1)}_\text{spect.}=
-\frac{if_\pi^2f_BC_F}{4 N_c^2}4\pi\alpha_s
\int_0^1 dxdyd\xi\,\Phi_{B1}(\xi)\phi_\pi(x)\phi_\pi(y) 
\frac{1}{\xi\bar{x}\bar{y}}.
\label{LOresult}
\end{equation}
Following (\ref{factform}) and the conventions of \cite{Beneke:2005vv} we
write our amplitude in the form:
\begin{equation}
A_{\text{spect.} i}=-im_B^2
\int_0^1dx dy d\xi \, T_i^\text{II}(x,y,\xi)
f_B\Phi_{B1}(\xi)f_\pi\phi_\pi(x)
f_\pi\phi_\pi(y).
\label{LO4}
\end{equation}
where in the case of $\bar{B}\to\pi^+\pi^-$ we define
\begin{eqnarray}
A_{\text{spect.}1} &=& \langle \Op_2\rangle_\text{spect.}
\nonumber\\
A_{\text{spect.}2} &=& \langle \Op_1\rangle_\text{spect.}
\label{LO5}
\end{eqnarray}
and in the case $\bar{B}\to\pi^0\pi^0$ we define 
\begin{eqnarray}
A_{\text{spect.}1} &=& \langle \Op_1\rangle_\text{spect.}
\nonumber\\
A_{\text{spect.}2} &=& \langle \Op_2\rangle_\text{spect.}.
\label{LO6}
\end{eqnarray}
Because we use the NDR-scheme which preserves Fierz transformations
for $\Op_1$ and $\Op_2$, 
$T^\text{II}_i$ has the same form for both decay channels. From
(\ref{LOresult}) and (\ref{LO4}) we get:
\begin{eqnarray}
T_1^{\text{II}(1)} &=&
4\pi\alpha_s\frac{C_F}{4N_c^2}\frac{1}{\xi\bar{x}\bar{y}m_B^2}
\nonumber\\
T_2^{\text{II}(1)} &=& 0 \quad .
\label{LO7}
\end{eqnarray}

%% file: diffeq/diffeq.tex
In this section I will discuss the extraction of subleading powers of
Feynman integrals with the {\itshape method of differential
  equations} \cite{Remiddi:1997ny,Caffo:1998du,Gehrmann:1999as}. 
This method will prove
to be easy to implement in a computer algebra system. The idea to
obtain the analytic expansion of Feynman integrals by tracing them 
back to differential equations has first been proposed in 
\cite{Remiddi:1997ny}. 
This method, which is demonstrated in \cite{Remiddi:1997ny} by the one-loop
two-point integral and in \cite{Caffo:1998du} by the two-loop sunrise
diagram, uses differential equations with respect to the small or
large parameter, in which the integral has to be expanded. 

In contrast to \cite{Remiddi:1997ny,Caffo:1998du} 
I will discuss the case where setting the small parameter to zero 
gives rise to new divergences. In this case the initial condition 
is not given by the differential equation itself and also cannot be 
obtained by calculation of the simpler integral that is defined by 
setting the expansion parameter to zero.
It is not possible to give a general proof, but it seems to be a rule, 
that one needs the leading power as a ``boundary condition''.
An efficient way to calculate the leading power of Feynman integrals
is provided by the \emph{method of regions}
\cite{Gorishnii:1989dd,Beneke:1997zp,Smirnov:1990rz,Smirnov:2002pj}, 
whereas the subleading powers can be obtained from a differential equation. 
In the present section I will discuss which conditions the differential 
equation has to fulfil in order for this to work.

Although the examples I use below are taken from the present calculation, this
method is very general and can be used in any case in which the
expansion of Feynman integrals in small parameters is needed.

\subsection{Description of the method}
We start with a (scalar) integral of the form 
\begin{equation}
I(p_1,\ldots,p_n,m_1,\ldots,m_n)=
\intd \frac{1}{D_1\ldots D_n}
\label{l1}
\end{equation}
where the propagators are of the form $D_i=(k+p_i)^2-m_i^2$. We assume
that there is only one mass hierarchy, i.e.\ there are two masses $m\ll
M$ such that all of the momenta and masses $p_i$ and $m_i$ are of
$\mathcal{O}(m)$ or of $\mathcal{O}(M)$. We expand (\ref{l1}) in
$\frac{m}{M}$ by replacing all small momenta and masses by $p_i\to\lambda
p_i$ and expand in $\lambda$. After the expansion the bookkeeping 
parameter $\lambda$ can be set to $1$. 

We obtain a differential equation for $I$ by differentiating the 
integrand in (\ref{l1}) with respect to $\lambda$. This gives rise 
to new Feynman integrals with propagators
of the form $\frac{1}{D_i^2}$ and scalar products $k\cdot p_i$ in the
numerator. Those Feynman integrals, however, can be reduced to the
original integral and to simpler integrals (i.e.\ integrals that contain
less propagators in the denominator) by using integration by parts 
identities.

Finally we obtain for (\ref{l1}) a differential equation of the form
\begin{equation}
\frac{d}{d\lambda}I(\lambda)=h(\lambda)I(\lambda)+g(\lambda)
\label{l2}
\end{equation}
where $h(\lambda)$ contains only rational functions of $\lambda$ and
$g(\lambda)$ can be expressed by Feynman integrals with a reduced
number of propagators. It is easy to see that $h$ and $g$ are unique if 
and only if $I$ and the integrals contained in $g$ are master integrals 
with respect to IBP-identities, i.e.\ they cannot be reduced to simpler
integrals by IBP-identities. If $I(\lambda)$ is divergent in 
$\epsilon=\frac{4-d}{2}$, $I$, $h$ and $g$ have to be expanded in 
$\epsilon$:
\begin{eqnarray}
I &=& \sum_i I_i \epsilon^i
\nonumber\\
h &=& \sum_i h_i \epsilon^i
\nonumber\\
g &=& \sum_i g_i \epsilon^i.
\label{l3}
\end{eqnarray}
Plugging (\ref{l3}) into (\ref{l2}) gives a system of differential
equations for $I_i$, similar to (\ref{l2}). In the next paragraph we
will consider an example for this case.

First let us assume that $h(\lambda)$ and $g(\lambda)$ have the
following asymptotic behaviour in $\lambda$:
\begin{eqnarray}
  h(\lambda) &=& h^{(0)}+\lambda h^{(1)}+\ldots\nonumber\\
  g(\lambda) &=& \sum_j \lambda^j g^{(j)}(\ln\lambda) 
\label{l4}
\end{eqnarray}
i.e.\ $h$ starts at $\lambda^0$, and we allow that $g$ starts at a negative
power of $\lambda$. We count $\ln\lambda$ as $\mathcal{O}(\lambda^0)$
so the $g^{(j)}$ may depend on $\ln\lambda$. This dependence, however, has
to be such that 
\begin{equation}
\lim_{\lambda\to 0}\lambda g^{(j)}(\ln\lambda)=0.
\label{l4.1}
\end{equation}
The condition (\ref{l4.1}) is fulfilled, if the $g^{(j)}$ are of the
form of a \emph{finite} sum
\begin{equation}
\sum_{n=n_0}^m a_n \ln^n\lambda.
\label{l4.2}
\end{equation}
The limit $m\to\infty$ however can spoil the expansion
(\ref{l4}). E.g.\ $e^{-\ln\lambda}=\frac{1}{\lambda}$ so the condition
(\ref{l4.1}) is not fulfilled, which is due to the fact that we must
not change the order of the limits $\lambda\to 0$ and $m\to\infty$.

Further we assume that also
$I(\lambda)$ starts at $\lambda^0$
\begin{equation}
  I(\lambda)= I^{(0)}(\ln\lambda)+\lambda I^{(1)}(\ln\lambda)
+\ldots
\label{l5}
\end{equation}
and plug this into (\ref{l2}) such that  we obtain an equation which gives 
$I^{(i)}$ recursively:
\begin{equation}
  \lambda^i I^{(i)}=\int_0^\lambda d\lambda^\prime {\lambda^\prime}^{i-1}
  \left(\sum_{j=0}^{i-1}
  h^{(j)}I^{(i-1-j)}(\ln\lambda^\prime)+g^{(i-1)}(\ln\lambda^\prime)\right).
\label{l6}
\end{equation}
I want to stress that, because $h$ starts at
$\mathcal{O}(\lambda^0)$, (\ref{l6}) is a recurrence relation, i.e.\
$I^{(j)}$ does not mix into $I^{(i)}$ if $j\ge i$.
As the integral is only well defined if $i\ge 1$,
we need the leading power $I^{(0)}$ as ``boundary condition''
and (\ref{l6}) will give us all the higher powers in $\lambda$. It is
easy to implement (\ref{l6}) in a computer algebra system, because we
just need the integration of polynomials and finite powers of logarithms.

A modification is needed if $h$ starts at $\lambda^{-1}$ i.e.\
\begin{equation}
  h=-\frac{n}{\lambda}+h^{(0)}+\ldots.
\label{l7}
\end{equation}
By replacing $\bar{I}\equiv \lambda^n I$ we obtain the differential
equation
\begin{equation}
  \frac{d}{d\lambda}\bar{I}=\left(\frac{n}{\lambda}+h\right)\bar{I}
  +\lambda^n g
\label{l8}
\end{equation}
which is similar to (\ref{l2}) and leads to
\begin{equation}
  \lambda^{i+n} I^{(i)}=\int_0^\lambda d\lambda^\prime {\lambda^\prime}^{i+n-1}
  \left(\sum_{j=0}^{i+n-1}
  h^{(j)}I^{(i-1-j)}(\ln\lambda^\prime)+g^{(i-1)}(\ln\lambda^\prime)\right),
\label{l9}
\end{equation}
which is valid for $i\ge 1-n$. So, if $I$ starts at
$\mathcal{O}(\lambda^{-n})$, the subleading powers result from the
leading power.  

\subsection{Examples}
We start with a pedagogic example:
\begin{ex}
\begin{equation}
I=\intd\frac{1}{k^2(k^2-\lambda)(k^2-1)}
\label{p1}
\end{equation}
\end{ex}
where $\lambda\ll 1$. The exact expression for this integral is given
by:
\begin{equation}
I=\frac{i}{(4\pi)^2}\frac{\ln\lambda}{1-\lambda}=
\frac{i}{(4\pi)^2}\ln\lambda(1+\lambda+\lambda^2+\ldots).
\label{p2}
\end{equation}
We see that $I$ diverges for $\lambda\to 0$. As described e.g.\ in 
\cite{Smirnov:2002pj} we can
obtain the leading power by expanding the integrand in the regions
$k\sim\sqrt{\lambda}$ and $k\sim 1$. This leads in the first region to
\begin{equation}
\intd \frac{-1}{k^2(k^2-\lambda)}=
-\frac{i}{(4\pi)^{2-\epsilon}}\Gamma(1+\epsilon)
\left(\frac{1}{\epsilon}+1-\ln\lambda\right)
\label{p3}
\end{equation}
and in the second region to
\begin{equation}
\intd \frac{1}{k^4(k^2-1)}=
\frac{i}{(4\pi)^{2-\epsilon}}\Gamma(1+\epsilon)
\left(\frac{1}{\epsilon}+1\right)
\label{p4}
\end{equation}
such that we finally obtain 
\begin{equation}
I^{(0)}(\ln\lambda)=\frac{i}{(4\pi)^2}\ln\lambda.
\label{p5}
\end{equation}
This is the result we obtain from the leading power of (\ref{p2}).
We write the derivative of $I$ with respect to $\lambda$ in the
following form:
\begin{equation}
\frac{d}{d\lambda}I=\frac{1}{1-\lambda}
\left[I-\intd\frac{1}{k^2(k^2-\lambda)^2}\right].
\label{p6}
\end{equation}
We obtained the right hand side of (\ref{p6}) by decomposing
$\frac{d}{d\lambda}I$ into partial fractions. Of course this
decomposition is not unique which is due to the fact that $I$ itself
is not a master integral but can be further simplified by partial
fractioning. From (\ref{p6}) and (\ref{l2}) we get:
\begin{eqnarray}
h &=& \frac{1}{1-\lambda}=1+\lambda+\lambda^2+\ldots
\nonumber\\
g &=& \frac{i}{(4\pi)^2}\frac{1}{\lambda(1-\lambda)}
=  \frac{i}{(4\pi)^2}\left(\lambda^{-1}+1+\lambda+\ldots \right)
\label{p7}
\end{eqnarray}
such that the coefficients in the expansion in $\lambda$ according to
(\ref{l4}) do not depend on the power label $(k)$:
\begin{equation}
h^{(k)}=1\quad\text{and}\quad g^{(k)}=\frac{i}{(4\pi)^2}.
\label{p8}
\end{equation}
We obtain for the recurrence relation (\ref{l6}):
\begin{equation}
I^{(k)}=\frac{1}{\lambda^k}\int_0^\lambda d\lambda^\prime\,
{\lambda^\prime}^{k-1}
\left(\sum_{j=0}^{k-1}I^{(k-1-j)}(\ln\lambda^\prime)
+\frac{i}{(4\pi)^2}\right).
\label{p9}
\end{equation}
Using the initial value (\ref{p5}) it is easy to prove by induction
\begin{equation}
I^{(k)}(\ln\lambda)=\frac{i}{(4\pi)^2}\ln\lambda
\quad\forall k\ge 0.
\label{p10}
\end{equation}
This result coincides with (\ref{p2}).

The first nontrivial example, we want to consider, 
is the following three-point integral:
\begin{ex}
\label{example2}
\begin{equation}
I=\intd \frac{1}{k^2(k+un_-+l)^2(k+n_++n_-)^2}.
\label{intex2}
\end{equation}
\end{ex}
Here $n_+$ and $n_-$ are collinear Lorentz vectors, which fulfil
$n_+^2=n_-^2=0$ and $n_+\cdot n_-=\frac{1}{2}$, $u$ is a real number
between $0$ and $1$ and $l$ is a Lorentz vector with $l^2=0$ and
$l^\mu\ll 1$. Furthermore we define 
\begin{equation}
\xi=2l\cdot n_+\quad\text{and}\quad \theta=2l\cdot n_-.
\end{equation}
We expand $I$ in $l$, so we make the replacement
$l\to\lambda l$ and differentiate $I$ with respect to $\lambda$. The integral
is not divergent in $\epsilon$ such that we obtain a
differential equation of the form (\ref{l2}) where the Taylor series
of $h(\lambda)$ starts at $\lambda^0$ as in (\ref{l4}). 
In $g(\lambda)$ only two-point integrals occur, which are easy to
calculate. I do not want to give the explicit expressions for $h$ and
$g$ because they are complicated, their exact form is not needed to
understand this example and they can be handled by a computer algebra
system. Because the leading power of $I$ is of
$\mathcal{O}(\lambda^0)$, (\ref{l6}) gives all of the subleading
powers.

We obtain the leading power as follows: First we have to identify the
regions, which contribute at leading power. If we decompose $k$ into 
\begin{equation}
k^\mu=2 k\cdot n_+n_-^\mu+2 k\cdot n_-n_+^\mu+k_\perp^\mu
\label{l10.1}
\end{equation}
we note that the only regions, which remain at leading power, are
the hard region $k^\mu\sim 1$ and the hard-collinear region 
\begin{eqnarray}
k\cdot n_+ &\sim& 1 \nonumber\\
k\cdot n_- &\sim& \lambda \nonumber\\
k_\perp^\mu &\sim& \sqrt{\lambda}.
\label{l10.2}
\end{eqnarray}
The soft region $k^\mu\sim\lambda$ leads at leading power to a
scaleless integral, which vanishes in dimensional regularisation.
In the hard region we expand the integrand to
\begin{equation}
\frac{1}{k^2(k+un_-)^2(k+n_++n_-)^2}.
\label{l10.3}
\end{equation}
By introducing a convenient Feynman parametrisation we obtain for the
$(4-2\epsilon)$-dimensional integral over (\ref{l10.3}):
\begin{equation}
\frac{i}{(4\pi)^{2-\epsilon}}\Gamma(1+\epsilon)\exp(i\pi\epsilon)
\frac{1}{u}
\left(
\frac{\ln(1-u)}{\epsilon}-\frac{1}{2}\ln^2(1-u)
\right).
\label{l10.4}
\end{equation}
In the hard-collinear region we expand the integrand to
\begin{equation}
\frac{1}{k^2(k+un_-+\theta n_+)^2(2k\cdot n_++1)}.
\label{l10.5}
\end{equation}
The integral over (\ref{l10.5}) gives:
\begin{equation}
\begin{split}
\frac{i}{(4\pi)^{2-\epsilon}}\Gamma(1+\epsilon)\exp(i\pi\epsilon)
\frac{1}{u}
\bigg(&
\frac{-\ln(1-u)}{\epsilon}
+2\li(u)
+\frac{1}{2}\ln^2(1-u)\\
&+\ln u\ln(1-u)+
\ln(1-u)\ln\theta
\bigg).
\end{split}
\label{l10.6}
\end{equation}
Adding (\ref{l10.4}) and (\ref{l10.6}) together we get the leading
power of (\ref{intex2}):
\begin{equation}
I^{(0)}=
\frac{i}{(4\pi)^2}\frac{1}{u}
\left(
2\li(u)+\ln u\ln(1-u)+\ln(1-u)\ln\theta
\right).
\label{l10.7}
\end{equation}
By plugging (\ref{l10.7}) into (\ref{l6}) we obtain
$I$ at $\mathcal{O}(\lambda)$:
\begin{equation}
\begin{split}
&I^{(1)}=\\
&\begin{split}
\frac{i}{(4\pi)^2}\frac{1}{u}\bigg[&
\theta\bigg(
-2+\ln u +\frac{\ln(1-u)\ln\theta}{u}+\frac{\ln(1-u)\ln u}{u}+
\ln\xi+\frac{2\li(u)}{u}
\bigg)-\\
&\begin{split}
\xi\bigg(&\frac{\ln u}{1-u}+2\frac{\ln(1-u)}{u}+
\frac{\ln(1-u)\ln\theta}{u}+
\frac{\ln(1-u)\ln u}{u}+\\
&\frac{\ln\xi}{1-u}+
\frac{2\li(u)}{u}
\bigg)
\bigg].
\end{split}
\end{split}
\end{split}
\label{l10.8}
\end{equation}

Now we want to consider the following four-point integral
\begin{ex}
\label{example1}
\begin{equation}
I=\intd\frac{1}{k^2(k+n_-)^2(k+l-n_+)^2(k+l-un_+)^2},
\label{intex1}
\end{equation}
\end{ex}
where we used the same variables, which were introduced in
(\ref{intex2}). This example is very special, because in this case our
method will allow us to obtain not only the subleading but also 
the leading power in $l$. $I$ is divergent in $\epsilon$ such that we
obtain after the expansion (\ref{l3}) a system of differential
equations of the following form:
\begin{eqnarray}
\frac{d}{d\lambda}I_{-1}&=&h_0I_{-1}+g_{-1}
\nonumber\\
\frac{d}{d\lambda}I_{0}&=&h_0I_{0}+h_1I_{-1}+g_{0}.
\label{intex1.2}
\end{eqnarray}
It turns out that in our example $h$ takes the simple form
\begin{equation}
h=-\frac{2+2\epsilon}{\lambda}
\end{equation}
such that analogously to (\ref{l8}) we can transform (\ref{intex1.2}) into
\begin{eqnarray}
\frac{d}{d\lambda}(\lambda^2 I_{-1})&=&\lambda^2g_{-1}
\nonumber\\
\frac{d}{d\lambda}(\lambda^2I_{0})&=&-2\lambda I_{-1}+\lambda^2g_{0}.
\label{intex1.3}
\end{eqnarray} 
This system of differential equations can easily be integrated to:
\begin{eqnarray}
I^{(i)}_{-1} &=& \frac{1}{\lambda^{i+2}}
\int_0^\lambda d\lambda^\prime\,{\lambda^\prime}^{i+1}g_{-1}^{(i-1)}
\nonumber\\
I^{(i)}_{0} &=& \frac{1}{\lambda^{i+2}}
\int_0^\lambda d\lambda^\prime\,{\lambda^\prime}^{i+1}
\left(-2I_{-1}^{(i)}+g_{0}^{(i-1)}\right)
\label{intex1.4}
\end{eqnarray}
where the superscript $(i)$ denotes the order in $\lambda$ as in
(\ref{l4}) and (\ref{l5}). 
Both $I_{-1}$ and $I_0$ start at $\mathcal{O}(\lambda^{-1})$. Because
(\ref{intex1.4}) is valid for $i\ge -1$, it gives us the leading power
expression, which reads:
\begin{equation}
I^{(-1)}=\frac{i}{(4\pi)^{2-\epsilon}}\Gamma(1+\epsilon)
\frac{2}{u\xi}
\left(
\frac{1}{\epsilon}-1-\frac{\ln u}{1-u}-\ln\xi
\right)
\label{intex1.5}
\end{equation}
where $\xi=2l\cdot n_+$ as in the example above. The exact expression
for (\ref{intex1}) can be obtained from \cite{Duplancic:2000sk}. Thereby
(\ref{intex1.5}) can be tested.

In the last paragraph I want to return to Example \ref{example2}.
I will show how we can use
differential equations to prove that the integral (\ref{intex2}) 
depends in leading power only on the soft kinematical variable 
$\theta=2l\cdot n_-$ and not on $\xi=2l\cdot n_+$.
We need derivatives of the integral with respect
to $\xi$ and $\theta$, which we have to express through derivatives with
respect to $l^\mu$. These derivatives can be applied directly to the
integrand, whose dependence on $l^\mu$ is obvious. We start from the
following equations:
\begin{eqnarray}
  n_+^\mu\frac{\partial}{\partial l^\mu} I &=&
  \frac{\partial}{\partial \theta}I+
  \xi \frac{\partial}{\partial l^2}I\nonumber\\
   n_-^\mu\frac{\partial}{\partial l^\mu} I &=&
  \frac{\partial}{\partial \xi}I+
  \theta \frac{\partial}{\partial l^2}I\label{ldiff1}\\
   l^\mu\frac{\partial}{\partial l^\mu} I &=&
  \xi \frac{\partial}{\partial \xi}I+
  \theta \frac{\partial}{\partial \theta}I+
  2l^2 \frac{\partial}{\partial l^2}I\nonumber
 \end{eqnarray}
which lead to
\begin{eqnarray}
 \xi \frac{\partial}{\partial \xi}I &=&
 \frac{1}{2}(-\theta n_+^\mu+\xi n_-^\mu+l^\mu)
 \frac{\partial}{\partial l^\mu} I\nonumber\\
 \theta \frac{\partial}{\partial \theta}I &=&
 \frac{1}{2}(\theta n_+^\mu-\xi n_-^\mu+l^\mu)
 \frac{\partial}{\partial l^\mu}I.\label{ldiff2}
\end{eqnarray}
where we have set $l^2=0$ in (\ref{ldiff2}). 
Using (\ref{ldiff2}) we can show that in leading power
(\ref{intex2}) depends only on $\theta$ and not
on $\xi$. So we can simplify the calculation of the leading power by
making the replacement $l^\mu\to\theta n_+^\mu$. The
proof goes as follows: From (\ref{l5}) we see that the statement
``$I^{(0)}$ does not depend on $\xi$'' is equivalent to 
\begin{equation}
  \xi\frac{\partial}{\partial \xi}I(\xi\lambda,\theta\lambda)=
  \mathcal{O}(\lambda).
\label{diffproof}
\end{equation}
Using the first equation of (\ref{ldiff2}) we get
\begin{equation}
   \xi\frac{\partial}{\partial \xi}I(\xi\lambda,\theta\lambda)=
   \mathcal{O}(\lambda) I(\xi\lambda,\theta\lambda)+\mathcal{O}(\lambda).
\end{equation}
Because we know (e.g.\ from power counting) that
$I(\xi\lambda,\theta\lambda)$ starts at $\lambda^0$, (\ref{diffproof})
is proven.

%% file: diagrams/diagrams_farxiv.tex
\subsection{Evaluation of the Feynman diagrams \label{eval}}
\begin{figure}
\begin{center}
\resizebox{0.8\textwidth}{!}{\includegraphics{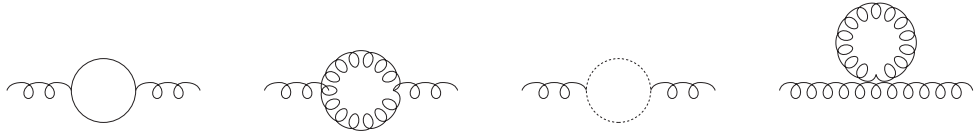}}
\end{center}
\caption{Gluon self energy}
\label{ag}
\end{figure}
\begin{figure}
\begin{center}
\resizebox{.8\textwidth}{!}{\includegraphics{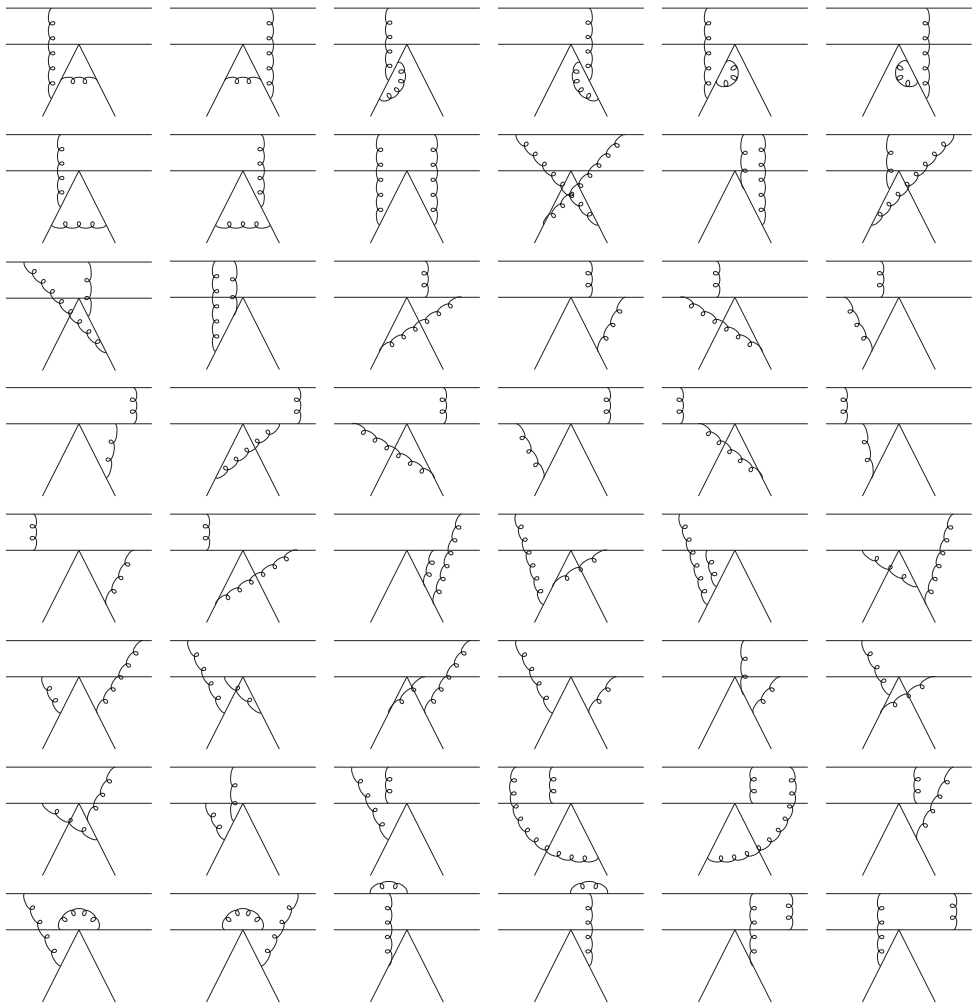}}
\end{center}
\caption{Abelian diagrams}
\label{abelian}
\end{figure}
\begin{figure}
\begin{center}
\resizebox{.8\textwidth}{!}{\includegraphics{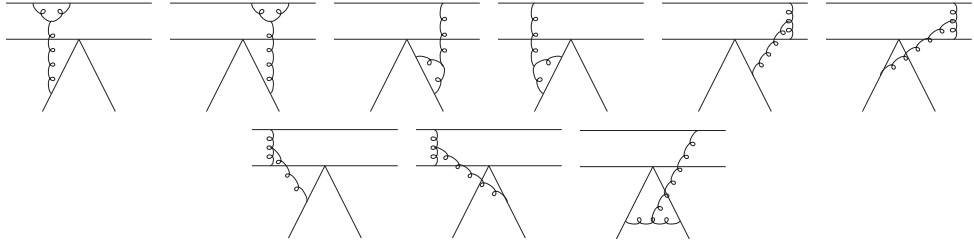}}
\end{center}
\caption{Nonabelian diagrams}
\label{nonabelian}
\end{figure}
The diagrams that contribute to $T^\text{II}_1$ at NLO are listed in
fig.~\ref{ag} - \ref{nonabelian}.
The diagrams are evaluated in leading power in $\lqcd/m_b$. We do not
use an effective theory like SCET but evaluate the diagrams in full
QCD using free external quark states, to which we assign the momenta
given in fig.~\ref{basic}. The Feynman integrals have to be evaluated
in leading power in $\lqcd/m_b$.
After reducing the number of Feynman integrals by integration by parts
(IBP) identities \cite{Chetyrkin:1981qh,Tkachov:1981wb}, we get the 
leading power of the master integrals using the
method of regions (see e.g.\ \cite{Smirnov:2002pj}). In many cases the
method of regions allows for a further
reduction of master integrals. E.g.\ the first diagrams in the second
line of fig.~\ref{nonabelian} comes with the scalar integral
\begin{equation}
\intd\frac{1}{k^2(k+\bar{x}p-l)^2(k+\bar{x}p+\bar{y}q-l)^2((k+p+q-l)^2-m_b^2)}.
\label{b67.1}
\end{equation}
This integral can be calculated in leading power by setting 
$\theta=0$, i.e.\ we make
the replacement $l^\mu\to\xi q^\mu$ where $\xi$ and $\theta$ are defined 
in (\ref{LO1}). This can be seen as follows: 
Counting soft momenta as $\mathcal{O}(\lambda)$ and hard momenta as
$\mathcal{O}(m_b)$ 
the regions of space where (\ref{b67.1}) gives a leading power
contribution are 
\begin{equation}
\begin{split}
&k^\mu \sim m_b\\
&k^\mu \sim \lambda \\
&k\cdot p \sim \lambda \quad
k^\mu_\perp \sim \sqrt{\lambda} \quad
k\cdot q \sim m_b.
\nonumber
\end{split}
\end{equation}
In these regions $l^\mu$ occurs only in the combination $l\cdot p$. 
So we can make the replacement $l^\mu\to\xi q^\mu$. It can be easily
seen that this replacement allows to reduce (\ref{b67.1}) to
three-point functions by decomposing the integrand into partial fractions.
Alternatively one can use the exact expression (\ref{4p6}) for the 
four-point integral 
with one massive propagator line, which is given in 
appendix~\ref{4pointmass}. After taking the leading power it can
easily be seen that we get the same result as by just making the 
replacement $l^\mu\to\xi q^\mu$.

Tensor integrals of the form $\intd (k^\mu,k^\mu
k^\nu,...)/\text{Denominator}$ can be reduced to scalar integrals
using the methods of \cite{Passarino:1978jh}. This reduction may make
it necessary to calculate subleading powers of  Feynman integrals to
obtain some diagrams in leading power. Furthermore, subleading powers
of Feynman
integrals are necessary in some diagrams, where the leading power vanishes
because of the equations of motion. The methods of the last section,
however, allow for an extraction of the subleading powers once the
leading powers have been calculated.
 
It is instructive and helps to avoid mistakes 
to obtain the Feynman integrals in two independent ways. 
Instead of using arguments depending on power counting we can obtain
the power expansion of the integrals by calculating the exact
integrals and expanding the results.
The exact expressions of massless four-point integrals
are given in \cite{Duplancic:2000sk}. Quite general expressions for
four-point integrals with one massive propagator are given in appendix
\ref{4pointmass}. All of the integrals that were used for the present
calculation have passed this independent test.

%% file: wavefunctions/wf.tex
\subsection{Wavefunction contributions}
\label{wf}
\subsubsection{General remarks}
\begin{figure}
\begin{center}
\resizebox{0.5\textwidth}{!}{\includegraphics{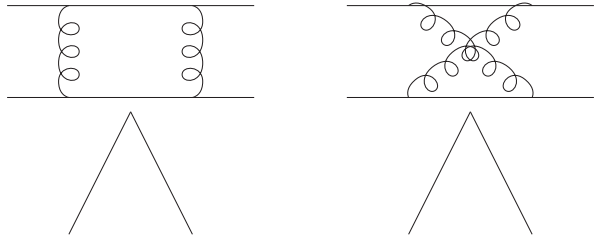}}
\end{center}
\caption{Example for diagrams which obviously belong to the form factor}
\label{ff2}
\end{figure}

It has already been demonstrated in section \ref{HspLO} how in principle we 
can extract the scattering kernel $T^\text{II}$ of (\ref{factform}) 
from the amplitude if we know the wave functions. 
$T^\text{II}$ does not depend on the hadronic
physics and on the form of the wave function $\phi_\pi$ and $\phi_B$ in
particular, so we can get $T^\text{II}$ by calculating the matrix
elements of the effective operators between free quark states carrying
the momenta shown in fig.~\ref{basic} on page \pageref{basic}. Because
we calculate $T^\text{II}$ in NLO we need unlike as in section \ref{HspLO}
the wave functions up to NLO. Let us write the second term of
(\ref{factform}) in the following formal way:
\begin{equation}
\mathcal{A}_\text{spect.}=
\phi_\pi\otimes\phi_\pi\otimes\phi_B\otimes T^\text{II}.
\label{wf1}
\end{equation}
All of the objects arising in (\ref{wf1}) have their perturbative
series in $\alpha_s$, so (\ref{wf1}) becomes
\begin{eqnarray}
\mathcal{A}_\text{spect.}^{(1)} &=&
\phi_\pi^{(0)}\otimes\phi_\pi^{(0)}\otimes\phi_B^{(0)}\otimes
T^{\text{II}(1)}
\label{wf2}\\
\mathcal{A}_\text{spect.}^{(2)} &=&
\phi_\pi^{(1)}\otimes\phi_\pi^{(0)}\otimes\phi_B^{(0)}\otimes
T^{\text{II}(1)}+
\phi_\pi^{(0)}\otimes\phi_\pi^{(1)}\otimes\phi_B^{(0)}\otimes
T^{\text{II}(1)}+
\nonumber\\
&&\phi_\pi^{(0)}\otimes\phi_\pi^{(0)}\otimes\phi_B^{(1)}\otimes
T^{\text{II}(1)}+
\phi_\pi^{(0)}\otimes\phi_\pi^{(0)}\otimes\phi_B^{(0)}\otimes
T^{\text{II}(2)}
\nonumber\\
&\vdots&
\nonumber
\end{eqnarray}
where the superscript $(i)$ denotes the order\footnote{
Note that the hard spectator scattering kernel starts at
$\mathcal{O}(\alpha_s)$. So we call $T^{\text{II}(1)}$ the LO and 
$T^{\text{II}(2)}$ the NLO.} in $\alpha_s$. 
In order to get $T^{\text{II}(2)}$ we have to calculate
$\mathcal{A}_\text{spect.}^{(2)}$, $\phi_\pi^{(1)}$ and $\phi_B^{(1)}$
for our final states. Then $T^{\text{II}(2)}$ is given by
\begin{eqnarray}
\lefteqn{\phi_\pi^{(0)}\otimes\phi_\pi^{(0)}\otimes\phi_B^{(0)}\otimes
T^{\text{II}(2)}=}
\label{wf3}\\
&&\mathcal{A}_\text{spect.}^{(2)}-
\phi_\pi^{(1)}\otimes\phi_\pi^{(0)}\otimes\phi_B^{(0)}\otimes
T^{\text{II}(1)}-
\phi_\pi^{(0)}\otimes\phi_\pi^{(1)}\otimes\phi_B^{(0)}\otimes
T^{\text{II}(1)}-
\nonumber\\
&&\phi_\pi^{(0)}\otimes\phi_\pi^{(0)}\otimes\phi_B^{(1)}\otimes
T^{\text{II}(1)}
\nonumber
\end{eqnarray}
At this point a subtlety occurs. Let us have a closer look to the
factorization formula (\ref{factform}). By calculating the first 
order in $\alpha_s$ of the partonic form factor $F^{B\to\pi,(1)}$,
which is defined by free quark states instead of hadronic external
states, we see that it can be written in the form
\begin{equation} 
F^{B\to\pi,(1)} = \phi_\pi^{(0)}\otimes\phi_B^{(0)}\otimes
T^{(1)}_\text{formfact.}.
\label{wf4}
\end{equation}
But $T^{(1)}_\text{formfact.}$ is not part of $T^{\text{II}}$. 
So we have to modify (\ref{wf3})
insofar as we have to subtract the right hand side of (\ref{wf4}) from
the right hand side of (\ref{wf3}):
\begin{eqnarray}
\lefteqn{\phi_\pi^{(0)}\otimes\phi_\pi^{(0)}\otimes\phi_B^{(0)}\otimes
T^{\text{II}(2)}=}
\label{wf5}\\
&&\mathcal{A}_\text{spect.}^{(2)}-
\phi_\pi^{(1)}\otimes\phi_\pi^{(0)}\otimes\phi_B^{(0)}\otimes
T^{\text{II}(1)}-
\phi_\pi^{(0)}\otimes\phi_\pi^{(1)}\otimes\phi_B^{(0)}\otimes
T^{\text{II}(1)}-
\nonumber\\
&&\phi_\pi^{(0)}\otimes\phi_\pi^{(0)}\otimes\phi_B^{(1)}\otimes
T^{\text{II}(1)}-
\phi_\pi^{(0)}\otimes\phi_\pi^{(0)}\otimes\phi_B^{(0)}\otimes
T^{(1)}_\text{formfact.}\otimes T^{\text{I}(1)}
\nonumber
\end{eqnarray}
In (\ref{wf5}) we did not include the term 
$F^{B\to\pi,(2)}\otimes T^{\text{I}(0)}$, because it is obviously identical
with the diagrams where the gluons do not interact with the emitted
pion (e.g.\ those of fig.~\ref{ff2}).
Those diagrams where not considered in the last section. So we do not
have to consider them here.

\begin{figure}
\begin{center}
\resizebox{0.5\textwidth}{!}{\includegraphics{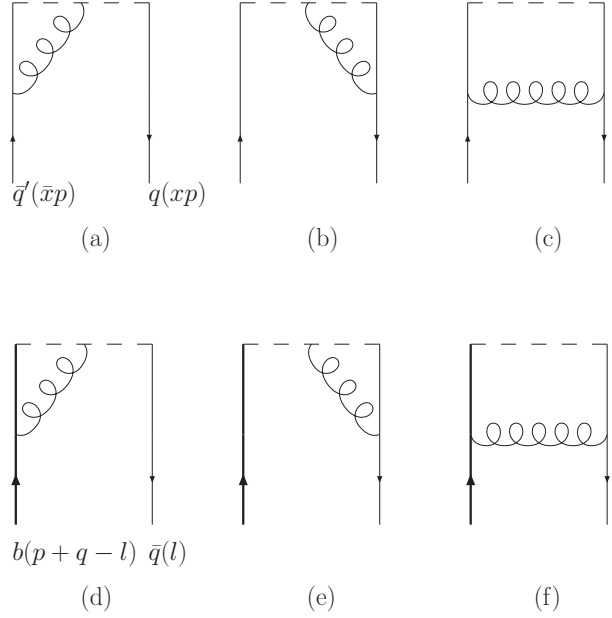}}
\end{center}
\caption{NLO contributions to the meson wave functions. The dashed
  line stands for the eikonal Wilson line which makes the wave
  functions gauge invariant.}
\label{wfpic1}
\end{figure}
The wave functions for free external quark states are given at LO by
(\ref{LO3}). At NLO there exist three possible contractions: The two
external quark states can be connected by a gluon propagator or one
of the external quarks can be connected to the eikonal Wilson line of
the wave function (fig.~\ref{wfpic1}).
The diagrams of fig.~\ref{wfpic1}(a),(b) and (c) give the order $\alpha_s$
of the ``pion wave function for free quarks'', i.e.\ we have
replaced the pion final state $\langle \pi(p)|$ in (\ref{mw1}) by the  
free quark state $\langle \bar{q}^\prime(\bar{x}p)q(xp)|$. The Fourier
transformed wave function $\phi_\pi^{(1)}(x^\prime)$ is defined
analogously to (\ref{LO3}). For the diagrams in fig.~\ref{wfpic1} (a), 
(b) and (c) respectively we get:
\begin{eqnarray}
\phi_{\pi\alpha\beta}^{\text{(a)},(1)}(x^\prime) &=&
8\pi^2 i \alpha_s C_F N_c \intd
\frac{\delta(x^\prime-x-\frac{k^+}{p^+})-\delta(x^\prime-x)}{k^2k^+}
\bar{q}_\beta(xp)
\left[\frac{1}{\sh{k}-\bar{x}\sh{p}}\gamma^+q^\prime(\bar{x}p)\right]_\alpha
\nonumber\\
\phi_{\pi\alpha\beta}^{\text{(b)},(1)}(x^\prime) &=&
8\pi^2 i \alpha_s C_F N_c \intd
\frac{\delta(x-x^\prime-\frac{k^+}{p^+})-\delta(x-x^\prime)}{k^2k^+}
\left[\bar{q}(xp)
\gamma^+\frac{1}{\sh{k}-x\sh{p}}
\right]_\beta q^\prime_\alpha(\bar{x}p)
\nonumber\\
\phi_{\pi\alpha\beta}^{\text{(c)},(1)}(x^\prime) &=&
8\pi^2 i \alpha_s C_F N_c \intd
\frac{\delta(x^\prime-x+\frac{k^+}{p^+})}{k^2}
\left[\bar{q}(xp)
\gamma^\tau\frac{1}{x\sh{p}-\sh{k}}
\right]_\beta 
\left[\frac{1}{\sh{k}+\bar{x}\sh{p}}\gamma_\tau
q^\prime(\bar{x}p)\right]_\alpha
\nonumber\\*
\label{wf6}
\end{eqnarray}

\subsubsection{Evanescent operators}
At NLO the convolution of the wave functions with the tree level
kernel $T^{\text{II},(1)}$ gives rise to new Dirac structures, which, however,
can in four dimensions be reduced to the tree level Dirac structures.
So we obtain the tree level Dirac structures plus further evanescent
structures, which vanish for $d=4$ but give finite contributions
if they are multiplied by UV-poles. We define our renormalisation 
scheme such that we subtract the UV-poles and these finite parts of the
evanescent structures. 

The tree level kernel (\ref{LOkernel}) contains two Dirac structures
where the second one is evanescent (after the the projection on the
wave functions). We write $T^{\text{II}(1)}$ in the following form:
\begin{equation}
T^{\text{II}(1)}(x,y,l^-)\equiv\frac{1}{\bar{x}l^-}
\gamma^\mu\tilde{\otimes}\gamma^\nu(1-\gamma_5)
\otimes\left(
\frac{2\sh{p}g_{\mu\nu}}{\bar{y}}-\frac{\sh{p}\gamma_\mu\gamma_\nu}{y\bar{y}}
\right)(1-\gamma_5)
\label{wf6.1}
\end{equation} 
where the symbol $\tilde{\otimes}$ stands for the ``wrong
contraction'' of the Dirac indices
i.e.\ the Dirac indices are given by
\begin{equation}
\left[\Gamma^1\tilde{\otimes}\Gamma^2\otimes\Gamma^3\right]
_{\alpha^\prime\alpha\beta^\prime\beta\gamma^\prime\gamma}
=
\Gamma^1_{\gamma^\prime\alpha}\Gamma^2_{\alpha^\prime\gamma}
\Gamma^3_{\beta^\prime\beta}
\label{wf6.2}
\end{equation}
as in (\ref{LOkernel}). The ``right contraction'' is defined by the
symbol $\otimes$ i.e.\ writing the Dirac indices explicitly
\begin{equation}
\left[\Gamma^1\otimes\Gamma^2\otimes\Gamma^3\right]
_{\alpha^\prime\alpha\beta^\prime\beta\gamma^\prime\gamma}
= \Gamma^1_{\alpha^\prime\alpha}\Gamma^2_{\gamma^\prime\gamma}
\Gamma^3_{\beta^\prime\beta}.
\label{wf6.3}
\end{equation}
In $d=4$ the wrong and the right contraction are related by Fierz
transformations. It is convenient and commonly used to define the
renormalised wave functions in terms of the right contraction,
i.e.\ to define $\phi_\pi^\text{ren.}$ by renormalising the
operator $\bar{q}(z)\gamma^\mu\gamma_5 q^\prime(0)$ instead of 
$\bar{q}(z)_\beta q^\prime(0)_\alpha$. This is why we define our
renormalisation scheme such that only the UV-finite part of the right
contraction operators remains: Using the notation of
(\ref{wf6.1}) -- (\ref{wf6.3}) we define the following operators:
\begin{eqnarray}
\Op_0(x,y,l^-) &\equiv& -\frac{1}{2l^-\bar{x}}
\gamma^\mu(1-\gamma_5)\otimes\gamma_\mu(1-\gamma_5)\otimes
\frac{2\sh{p}}{\bar{y}}(1-\gamma_5)
\label{wf6.4}\\
\Op_1(x,y,l^-) &\equiv& \frac{1}{l^-\bar{x}}
\gamma^\mu\tilde{\otimes}\gamma_\mu(1-\gamma_5)\otimes
\frac{2\sh{p}}{\bar{y}}(1-\gamma_5)
\label{wf6.5}\\
\Op_2(x,y,l^-) &\equiv& \frac{1}{l^-\bar{x}}
\gamma^\mu\tilde{\otimes}\gamma^\nu(1-\gamma_5)\otimes
\frac{-\sh{p}\gamma_\mu\gamma_\nu}{y\bar{y}}(1-\gamma_5).
\label{wf6.6}
\end{eqnarray}
The matrix elements of these operators are defined analogously to 
(\ref{LO3.1}):
\begin{equation}
\langle\Op_i\rangle\equiv
\int dx^\prime dy^\prime dl^{\prime-}\,\phi_{\pi\alpha\alpha^\prime}(x^\prime)
\phi_{\pi\beta\beta^\prime}(y^\prime)
\phi_{B\gamma\gamma^\prime}(l^{\prime-})
\Op_{i\,\alpha^\prime\alpha\beta^\prime\beta\gamma^\prime\gamma}
(x^\prime,y^\prime,l^{\prime-}).
\label{wf6.7}
\end{equation}
Note that $\langle\Op_1+\Op_2\rangle$ is just the convolution of the
tree level kernel (\ref{wf6.1}) with the wave functions. Furthermore
by using Fierz identities it is easy to prove that we have in four dimensions 
\begin{eqnarray}
\langle\Op_0\rangle &=& \langle\Op_1\rangle
\label{wf6.8}\\
\langle\Op_2\rangle &=& 0.
\label{wf6.9}
\end{eqnarray}
So we define the following evanescent operators:
\begin{eqnarray}
E_1&\equiv&\Op_2
\label{wf6.10}\\
E_2&\equiv&\Op_1-\Op_0
\label{wf6.11}\\
E_3&\equiv&\frac{1}{\bar{x}\bar{y}l^-}
\left(
\gamma^\mu\gamma^\nu\gamma^\rho\tilde{\otimes}
\gamma_\rho\gamma_\nu\gamma_\mu(1-\gamma_5)
+\frac{(2-d)^2}{2}
\gamma^\mu(1-\gamma_5)\otimes\gamma_\mu(1-\gamma_5)
\right)
\nonumber\\
&&\otimes
2\sh{p}(1-\gamma_5)
\label{wf6.12}\\
E_4&\equiv&\frac{1}{\bar{x}y\bar{y}l^-}
\gamma^\mu\gamma^\lambda\gamma^\tau\tilde{\otimes}
\gamma_\tau\gamma_\lambda\gamma^\nu(1-\gamma_5)\otimes
\sh{p}\gamma_\mu\gamma_\nu(1-\gamma_5)
\label{wf6.13}
\end{eqnarray}
where we have defined $E_3$ and $E_4$ for later convenience.
Using those operator definitions we define our renormalisation scheme
such that we subtract the UV-pole of $\langle\Op_0\rangle$ and 
the finite parts of
$\langle E_i\rangle$ i.e.\ terms of the form
$\frac{1}{\epsilon_\text{UV}}\langle E\rangle$, where $\langle
E\rangle$ is an arbitrary evanescent structure. 
It is important to note that we do not subtract IR-poles, because they  
depend not only on the operator but also on the external states the
operator is sandwiched in between. They have to vanish in (\ref{wf5}) 
such that the hard scattering kernel is finite. 
Finally we obtain the same result as if we had regularised the 
IR-divergences by small quark and gluon masses because the evanescent
structures vanish in $d=4$. The renormalisation scheme defined above
is the same scheme that was used in \cite{Beneke:2005vv}.

In the next step we will calculate the convolution integral of
$T^{\text{II},(1)}$ with the NLO wave functions given by
(\ref{wf6}), 
 i.e.\ we
have to calculate the renormalised matrix elements of $\Op_1+\Op_2$ at
NLO.

\subsubsection{Wave function of the emitted pion}
First we consider the renormalisation of the emitted pion wave function:
Because the contribution of the wave functions $\phi_\pi^{(a)}$ 
and $\phi_\pi^{(b)}$ does not change the Dirac structure of the
operators, we do not need to consider evanescent operators when we
calculate the diagrams of fig.~\ref{wfpic1}(a),(b). So for the emitted
pion wave function these diagrams give after renormalisation:
\begin{equation}
\begin{split}
\langle\Op_1^\text{ren.}+\Op_2^\text{ren.}
\rangle^{(1),\text{(a),(b)}}_\text{emitted}
=
\frac{2\alpha_s}{4\pi}C_F\bigg[&
\left(-\frac{1}{\epsilon_\text{IR}}+
2\ln\frac{\mu_\text{UV}}{\mu_\text{IR}}\right)
\frac{\ln\bar{y}+2y}{y}\langle\Op_1\rangle^{(0)}
\\
&-
\left(\frac{1}{\epsilon_\text{IR}}+
2\ln\mu_\text{IR}\right)
\left(2+\ln y+\ln\bar{y}\right)
\langle\Op_2\rangle^{(0)}
\bigg]
\label{wf7}
\end{split}
\end{equation}
where the LO matrix elements $\langle\Op_i\rangle^{(0)}$ can be
obtained from (\ref{LO2}). Note that we kept the IR-pole times
the evanescent matrix element $\langle\Op_2\rangle^{(0)}$. This is
needed for consistency because we also kept similar terms in the
QCD-calculation of $\mathcal{A}_\text{spect.}$. Furthermore it allows
us to show that all IR-divergences vanish.

The diagram in fig.~\ref{wfpic1}(c) mixes different Dirac structures. So
we have to include evanescent operators in the renormalisation. In the
case of the emitted pion wave function the operator $\Op_1$ does not
mix under renormalisation with the evanescent operator $E_1$
(\ref{wf6.10}). We obtain for the renormalised matrix element:
\begin{equation}
\langle\Op_1^\text{ren.}\rangle^{(1),\text{(c)}}_\text{emitted}
=
-\frac{2\alpha_s}{4\pi}C_F\frac{\bar{y}\ln\bar{y}}{y}
\left(-\frac{1}{\epsilon_\text{IR}}+
2\ln\frac{\mu_\text{UV}}{\mu_\text{IR}}\right)
\langle\Op_1\rangle^{(0)}.
\label{wf7.1}
\end{equation}
The matrix element of $E_1$ however has an overlap with $\Op_1$:
\begin{equation}
\begin{split}
\langle E_1\rangle^{(1),\text{(c)}}_\text{emitted}=
\frac{\alpha_s}{4\pi}C_F
\left(
\frac{1}{\epsilon_\text{UV}}-\frac{1}{\epsilon_\text{IR}}+
2\ln\frac{\mu_\text{UV}}{\mu_\text{IR}}
\right)
\bigg[&
\left(-2y\ln y-2\bar{y}\ln\bar{y}\right)
\langle E_1\rangle^{(0)}\\
&-4\epsilon\left(\ln y+\frac{\bar{y}\ln\bar{y}}{y}\right)
\langle\Op_1\rangle^{(0)}
\bigg].
\label{wf7.2}
\end{split}
\end{equation}
The renormalisation prescription tells us to subtract the UV-pole and
the UV-finite part of $\langle E_1\rangle$. So we obtain after
renormalisation:
\begin{equation}
\begin{split}
\langle E_1^\text{ren.}\rangle^{(1),\text{(c)}}_\text{emitted}=
\frac{\alpha_s}{4\pi}C_F
\bigg[&
\left(
-\frac{1}{\epsilon_\text{IR}}+
2\ln\frac{\mu_\text{UV}}{\mu_\text{IR}}
\right)
\left(-2y\ln y-2\bar{y}\ln\bar{y}\right)
\langle E_1\rangle^{(0)}\\
&+4\left(\ln y+\frac{\bar{y}\ln\bar{y}}{y}\right)
\langle\Op_1\rangle^{(0)}
\bigg].
\label{wf7.3}
\end{split}
\end{equation} 
Note that the evanescent operator $E_1$ leads to a finite term $4(\ln
y+\frac{\bar{y}\ln\bar{y}}{y})\langle\Op_1\rangle^{(0)}$, which we
would have missed if we had just dropped the evanescent operators.

\subsubsection{Wave function of the recoiled pion}
In the next step we consider the NLO contribution of the recoiled pion
wave function. As in the case of the emitted pion the diagrams
fig.~\ref{wfpic1}(a),(b) do not lead to a mixing between the
operators. Therefore we get:
\begin{equation}
\begin{split}
\langle\Op_1^\text{ren.}+\Op_2^\text{ren.}
\rangle^{(1),\text{(a),(b)}}_\text{recoiled}=
\frac{\alpha_s}{4\pi}C_F\frac{2\ln\bar{x}+4x}{x}
\bigg[&
\left(-\frac{1}{\epsilon_\text{IR}}+
2\ln\frac{\mu_\text{UV}}{\mu_\text{IR}}\right)
\langle\Op_1\rangle^{(0)}\\
&-\left(\frac{1}{\epsilon_\text{IR}}+
2\ln\mu_\text{IR}\right)
\langle\Op_2\rangle^{(0)}
\bigg].
\end{split}
\label{wf7.4}
\end{equation}

Other than in the case of the emitted pion the operators $\Op_1$ and
$\Op_2$ mix the spinors of the recoiled pion and the
$B$-meson. Therefore we have to work in the operator basis of $\Op_0$
and the evanescent operators and define our renormalisation scheme
such that the finite parts of the matrix elements of the evanescent
operators vanish. The diagram fig.~\ref{wfpic1}(c) contributes to the
matrix element of the renormalised operator $\Op_0^\text{ren.}$:
\begin{equation}
\langle\Op_0^\text{ren.}\rangle^{(1),\text{(c)}}_\text{recoiled}
=2\frac{\alpha_s}{4\pi}C_F
\left(\frac{1}{\epsilon_\text{IR}}-
2\ln\frac{\mu_\text{UV}}{\mu_\text{IR}}
\right)
\frac{\bar{x}\ln\bar{x}}{x}
\langle\Op_0\rangle^{(0)}.
\label{wf7.5}
\end{equation}
In the case of the evanescent operators we keep the IR-pole:
\begin{eqnarray}
\langle E_1^\text{ren.}\rangle^{(1),\text{(c)}}_\text{recoiled}
&=&
\frac{1}{2}\frac{\alpha_s}{4\pi}C_F
\left(\frac{1}{\epsilon_\text{IR}}+2\ln\mu_\text{IR}\right)
\frac{\bar{x}\ln\bar{x}}{x}
\langle E_4 \rangle^{(0)}
\label{wf7.6}\\
\langle E_2^\text{ren.}\rangle^{(1),\text{(c)}}_\text{recoiled}
&=&
\frac{1}{4}\frac{\alpha_s}{4\pi}C_F
\left(\frac{1}{\epsilon_\text{IR}}+2\ln\mu_\text{IR}\right)
\frac{\bar{x}\ln\bar{x}}{x}
\langle E_3 \rangle^{(0)}
\label{wf7.8}
\end{eqnarray}
At the end of the day we obtain a contribution from diagram
fig.~\ref{wfpic1}(c):
\begin{equation}
\begin{split}
&\langle\Op_0^\text{ren.}+E_1^\text{ren.}+E_2^\text{ren.}
\rangle^{(1),\text{(c)}}_\text{recoiled}=
\\
&\quad
\begin{split}
\frac{1}{2}\frac{\alpha_s}{4\pi}C_F\frac{\bar{x}\ln\bar{x}}{x}
\bigg[&
\left(\frac{1}{\epsilon_\text{IR}}-2\ln\frac{\mu_\text{UV}}{\mu_\text{IR}}
\right)
\langle\frac{1}{\bar{x}l^-}\gamma^\mu\gamma^\lambda\gamma^\tau
\tilde{\otimes}\gamma_\tau\gamma_\lambda\gamma^\nu\otimes
\left(\frac{2\sh{p}g_{\mu\nu}}{\bar{y}}-
\frac{\sh{p}\gamma_\mu\gamma_\nu}{y\bar{y}}
\right)
\rangle^{(0)}
\\
&
+8\langle\Op_1\rangle^{(0)}
\bigg].
\end{split}
\end{split}
\label{wf7.9}
\end{equation}
The very complicated but also very explicit form, in which the above equation
was given, is rather convenient, because the QCD amplitude
$\mathcal{A}_\text{spect.}^{(2)}$ (see (\ref{wf1})-(\ref{wf3}))
comes with the same Dirac structure and cancels the IR-pole of (\ref{wf7.9}).

\subsubsection{Wave function of the $B$-meson}
The $\alpha_s$ corrections of the wave function of the $B$-meson are
given by the second row of fig.~\ref{wfpic1}. For the diagrams (d),
(e) and (f) respectively they read:
\begin{eqnarray}
\lefteqn{\phi_{B\alpha\beta}^{\text{(d)},(1)}(l^{\prime -}) =}
\label{wf8}\\
&&8\pi^2 i \alpha_s N_c C_F \intd
\frac{\delta(l^{\prime -}-l^--k^-)-\delta(l^{\prime -}-l^-)}{k^2k^-}
\left[\bar{q}(l)\gamma^-\frac{1}{\sh{k}+\sh{l}}\right]_\beta
b_\alpha(p+q-l)
\nonumber\\
\lefteqn{\phi_{B\alpha\beta}^{\text{(e)},(1)}(l^{\prime -}) =}
\nonumber\\*
&&8\pi^2 i \alpha_s N_c C_F \intd
\frac{\delta(l^{\prime -}-l^--k^-)-\delta(l^{\prime -}-l^-)}{k^2k^-}
\times
\nonumber\\*
&&\bar{q}_\beta(l)\left[
\frac{1}{\sh{k}-\sh{p}-\sh{q}+\sh{l}+m_b}\gamma^-b(p+q-l)
\right]_\alpha
\nonumber\\
\lefteqn{\phi_{B\alpha\beta}^{\text{(f)},(1)}(l^{\prime -}) =}
\nonumber\\
&&8\pi^2 i \alpha_s N_c C_F \intd
\frac{\delta(l^{\prime -}-l^--k^-)}{k^2}
\times
\nonumber\\
&&\left[\bar{q}(l)\gamma^\mu\frac{1}{\sh{k}+\sh{l}}\right]_\beta
\left[
\frac{1}{\sh{p}+\sh{q}-\sh{l}-\sh{k}-m_b}\gamma_\mu b(p+q-l)
\right]_\alpha
\nonumber
\end{eqnarray}

In the case of the $B$-meson only the diagrams in
fig.~\ref{wfpic1}(d),(e) give rise to UV-poles. Those diagrams however
do not lead to a mixing of $\Op_0$ and the evanescent operators and we
do not have to deal with evanescent operators.

First let us have a look at the convolution integral which belongs to
the diagram in fig.~\ref{wfpic1}(f):
\begin{eqnarray}
\lefteqn{\langle\Op_1+\Op_2\rangle^{(1),\text{(f)}}_B=}
\label{wf9}\\
&&(4\pi)^2i\alpha_s^2N_cC_F^2\frac{1}{\bar{x}}
\intd\frac{1}{2(k+l)\cdot p\,k^2}
\times
\nonumber\\
&&
\bar{q}_c
\gamma^\tau\frac{\sh{k}+\sh{l}}{(k+l)^2}\gamma^\mu
q_r^\prime
\bar{q}_r
\gamma^\nu(1-\gamma_5)
\frac{\sh{p}+\sh{q}-\sh{l}-\sh{k}+m_b}{k^2-2k\cdot(p+q-l)}
\gamma_\tau
b
\bar{q}_e
\left(
\frac{2\sh{p}}{\bar{y}}g_{\mu\nu}-
\frac{\sh{p}}{y\bar{y}}\gamma_\mu\gamma_\nu\right)(1-\gamma_5)
q_e^\prime,
\nonumber
\end{eqnarray}
where $q_c$, $q_r^{(\prime)}$ and $q_e^{(\prime)}$ are the spinors 
carrying the flavour quantum numbers of the light constituent
quark of the $B$-meson, the recoiled pion and the emitted pion
respectively.
\begin{figure}
\begin{center}
\resizebox{0.5\textwidth}{!}{\includegraphics{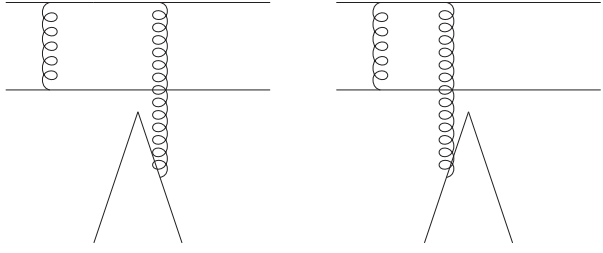}}
\end{center}
\caption{Two diagrams which correspond in leading power to the
  contribution of the $B$-meson wavefunction (\ref{wf9}).}
\label{wfpic2}
\end{figure}
In leading power (\ref{wf9}) is identical to the contribution of the
two diagrams shown in fig.~\ref{wfpic2}, which is given by:
\begin{equation}
\begin{split}
&-(4\pi)^2i\alpha_s^2N_cC_F^2 
\intd
\frac{1}{k^2(k+l-\bar{x}p)^2}
\\
&\quad\times
\bar{q}_c\gamma^\tau\frac{\sh{k}+\sh{l}}{(k+l)^2}\gamma^\mu 
q_r^\prime
\bar{q}_r
\gamma^\nu(1-\gamma_5)
\frac{\sh{p}+\sh{q}-\sh{l}-\sh{k}+m_b}{k^2-2k\cdot(p+q-l)}
\gamma_\tau
b
\\
&\quad
\bar{q}_e
\left(
\gamma_\mu
\frac{y\sh{q}+\bar{x}\sh{p}-\sh{k}-\sh{l}}{(yq+\bar{x}p-k-l)^2}
\gamma_\nu-
\gamma_\nu
\frac{\bar{y}\sh{q}+\bar{x}\sh{p}-\sh{k}-\sh{l}}{(\bar{y}q+\bar{x}p-k-l)^2}
\gamma_\mu\right)(1-\gamma_5)
q_e^\prime.
\end{split}
\label{wf10}
\end{equation}
In (\ref{wf10}) the leading power comes from the region where $k$ is
soft. In this region of space the integrand gets the form of the
integrand in (\ref{wf9}), so both contributions cancel. As we did not
include the diagrams of fig.~\ref{wfpic2} in the last section we can skip
the contribution of (\ref{wf9}) here. 

The remaining contributions are the diagrams in fig.~\ref{wfpic1}(d)
and (e). Together they read:
\begin{eqnarray}
\lefteqn{\langle\Op_1+\Op_2\rangle^{(1),\text{(d),(e)}}_B
=}
\label{wf11}\\
&&\alpha_s^2N_cC_F^2\frac{1}{\xi\bar{x}}
\bar{q}_c
\gamma^\mu
q_r^\prime\bar{q}_r
\gamma^\nu(1-\gamma_5)
b
\bar{q}_e
\left(
\frac{2\sh{p}}{\bar{y}}g_{\mu\nu}-
\frac{\sh{p}}{y\bar{y}}\gamma_\mu\gamma_\nu
\right)
q_e^\prime
\times
\nonumber\\
&&\left(
\left(\frac{1}{\epsilon_\text{UV}}+2\ln\frac{\mu_\text{UV}}{m_b}\right)
(4+2\ln\xi)-
2\left(\frac{1}{\epsilon_\text{IR}}+2\ln\frac{\mu_\text{IR}}{m_b}\right)
+4-\frac{2\pi^2}{3}-2\ln^2\xi\right).
\nonumber
\end{eqnarray}

\subsubsection{Form factor contribution}
\begin{figure}[t]
\begin{center}
\resizebox{0.5\textwidth}{!}{\includegraphics{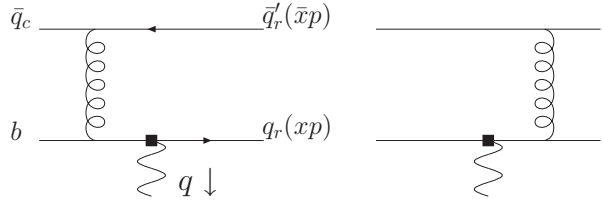}}
\end{center}
\caption{$\alpha_s$ contributions to the form factor}
\label{wfpic3}
\end{figure}
Finally we have to calculate the contribution of (\ref{wf4}). It is
given by
\begin{eqnarray}
\mathcal{A}_\text{formfact.}
&\equiv&
\phi_\pi^{(0)}\otimes\phi_\pi^{(0)}\otimes\phi_B^{(0)}\otimes
T^{(1)}_\text{formfact.}\otimes T^{\text{I}(1)}
\equiv
\label{wf12}\\
&&\frac{C_F\alpha_s}{4\pi}
f^{(1),\nu}\,\bar{q}_e(yq)\gamma_\nu(1-\gamma_5)q^\prime_e(\bar{y}q)
\,T^{(1)}(y).
\nonumber
\end{eqnarray}
The form factor $f^{(1),\nu}$ is the $\alpha_s$ correction of the
matrix element 
\begin{equation}
\label{wf13}
\langle\bar{q}_r^\prime(\bar{x}p)q_r(xp)|
\bar{q}_r\gamma^\nu(1-\gamma_5)b
|b(p+q-l)\bar{q}_c(l)\rangle,
\end{equation}
where $T^{(1)}(y)$ is given by
\cite{Beneke:2001ev}:
\begin{eqnarray}
T^{(1)}(y) &=&
-6\left(\frac{1}{\epsilon}+\lmu\right)-18+3\left(
\frac{1-2y}{\bar{y}}\ln y-i\pi\right)+
\label{wf14}\\
&&\left[
2\li(y)-\ln^2y+\frac{2\ln y}{\bar{y}}-(3+2i\pi)\ln y 
-(y\leftrightarrow\bar{y})
\right].
\nonumber
\end{eqnarray}
We get $f^{(1),\nu}$ by evaluating the diagrams in
fig.~\ref{wfpic3} and obtain finally:
\begin{eqnarray}
\mathcal{A}_\text{formfact.} &=& \alpha_s^2N_cC_F^2
\frac{1}{\bar{x}\xi}
\bar{q}_c
\gamma^\mu
q_r^\prime\bar{q}_r
\left(\gamma_\mu\frac{\sh{l}}{\xi}\gamma^\nu(1-\gamma_5)
-\gamma^\nu(1-\gamma_5)\frac{x\sh{p}+\sh{q}+1}{\bar{x}}\gamma_\mu\right)
b
\nonumber\\
&&\bar{q}_e\gamma_\nu(1-\gamma_5)q_e^\prime T^{(1)}(y).
\label{wf15}
\end{eqnarray}

%% file: result/result.tex
After the analysis of the last chapter we finally obtain the
$\mathcal{O}(\alpha_s^2)$ 
results for the hard spectator scattering kernels $T^\text{II}_{1,2}$ 
which are defined by (\ref{LO4}). Those expressions appear in convolution
integrals with wave functions, where $x$, $y$ and $\xi$ are the
integration variables as defined in (\ref{LO4}). The ultraviolet
divergences are renormalised in the $\overline{\text{MS}}$-scheme. The
infrared divergences drop out after subtracting the wave function
contributions from the amplitude.
The infrared finiteness together with the finiteness of the 
convolution integrals ensures that the framework of QCD-factorization 
works at this order in $\alpha_s$.  

The explicit $\mathcal{O}(\alpha_s^2)$ contributions for
$T^\text{II}_{1,2}$ read (see next page):
\vfill
\pagebreak[4]
\enlargethispage*{2cm}
\thispagestyle{empty}
\begin{eqnarray}
\input{result/T1re}
\end{eqnarray}
\begin{eqnarray}
\input{result/T1im}
\end{eqnarray}
\begin{eqnarray}
\input{result/T2re}
\end{eqnarray}
\pagebreak[4]
\begin{eqnarray}
\input{result/T2im}
\end{eqnarray}

The $\alpha_s^2$ corrections of the hard spectator interactions have
already been calculated in \cite{Beneke:2005vv,Kivel:2006xc}. However
both of these calculations have been performed in the framework of
SCET, while my result is a pure QCD calculation. In order to compare
(\ref{T1re})-(\ref{T2im}) to \cite{Beneke:2005vv,Kivel:2006xc} we have to
take into account the definition of $\lambda_B$. The SCET calculation
naturally uses the $\lambda_B$ defined by the HQET field for the
$b$-meson, while I define $\lambda_B$ by QCD-fields. Those two
definitons differ at $\mathcal{O}(\alpha_s)$, which has been discussed
in \cite{Pilipp:2007sb}. The difference in the logarithmic moments
of the $B$-meson wave function does not play a role, because these
moments occur first at NLO. Using the results of \cite{Pilipp:2007sb}
it is easy to figure out with the help of a computer algebra system, that
(\ref{T1re})-(\ref{T2im}) reproduce the results of  
\cite{Beneke:2005vv,Kivel:2006xc}.

%% file: result/T1re.tex
\text{Re}T^{\text{II}(2)}_1&=&-\frac{\alpha_s^2C_F}{4N_c^2m_B^2\xi}\times
\label{T1re}\\
&&\Bigg[C_N\bigg(-\frac{16 \ln  \xi }{3 \bar{x} \bar{y}}-\frac{16 \ln  \bar{x}}{3 \bar{x} \bar{y}}+\frac{40 \ln  \frac{\mu }{m_b}}{3 \bar{x} \bar{y}}+\frac{80}{9 \bar{x} \bar{y}}\bigg)\nonumber\\
&&\begin{split}
+C_F\bigg(
&\left(\frac{4 \ln  \xi }{\bar{x} \bar{y}}+\frac{4 \ln
    \bar{x}}{\bar{x} \bar{y}}+\frac{4 \ln  \bar{y}}{\bar{x}
    \bar{y}}+\frac{30}{\bar{x} \bar{y}}\right) 
\ln\frac{\mu }{m_b}
\\
&-\frac{\ln ^2\xi }{\bar{x} \bar{y}}
+\ln\xi \left(-\frac{2 \ln  x}{\bar{x}^2 \bar{y}}-
\frac{2 \ln  \bar{x}}{\bar{x} \bar{y}}-\frac{5}{\bar{x} \bar{y}}\right)
\\
&+\left(-\frac{2 \bar{x}^2}{(y-\bar{x})^3}-\frac{4
    \bar{x}}{(y-\bar{x})^2}-
    \frac{2}{y-\bar{x}}-\frac{2 x}{(y-x) \bar{x}}-\frac{2}{y \bar{x}^2}+
    \frac{2 (5 x-2)}{\bar{y} \bar{x}^2}\right) \li x
\\
&+\left(-\frac{2 \bar{x}^2}{(y-\bar{x})^3}-\frac{4 \bar{x}}{(y-\bar{x})^2}-\frac{2}{y-\bar{x}}+\frac{2 x}{(y-x) \bar{x}}-\frac{4}{\bar{x}}+\frac{2}{y \bar{x}^2}+\frac{4}{\bar{y} \bar{x}^2}\right) \li y
\\
&+\left(\frac{2 (x-2)}{\bar{x}^2 \bar{y}}+\frac{2}{\bar{x}}\right) \li(x y)
\\
&+\left(-\frac{2 \bar{x}^2}{(y-\bar{x})^3}-\frac{4 \bar{x}}{(y-\bar{x})^2}-\frac{2}{y-\bar{x}}\right) \li\left(-\frac{x y}{\bar{x}}\right)
\\
&+\left(\frac{2 x}{(y-x) \bar{x}}+\frac{2}{\bar{x} \bar{y}}\right) 
  \li\left(-\frac{y \bar{x}}{\bar{y}}\right)
+\left(-\frac{2}{\bar{x}}+\frac{2}{\bar{x}^2 \bar{y}}+
 \frac{2}{\bar{x}^2 y}\right) \li(x \bar{y})
\\
&+\left(-\frac{2 x}{(y-x) \bar{x}}-\frac{2}{\bar{x} \bar{y}}\right) 
\li\left(-\frac{x \bar{y}}{\bar{x}}\right)
\\
&+\left(\frac{2 \bar{x}^2}{(y-\bar{x})^3}+\frac{4\bar{x}}{(y-\bar{x})^2}+
\frac{2}{y-\bar{x}}\right) \li\left(-\frac{\bar{x} \bar{y}}{y}\right)
\\
&+\left(-\frac{2}{\bar{y} \bar{x}}-\frac{2}{\bar{x}}\right) \ln  x \ln  y
+\frac{2 (3 x-2) \ln  x \ln  \bar{x}}{\bar{x}^2 \bar{y}}
+\left(\frac{2}{\bar{x}}+\frac{2}{\bar{y} \bar{x}^2}\right)\ln x \ln\bar{y}
\\
&+\left(-\frac{2 \bar{x}^2}{(y-\bar{x})^3}-\frac{4\bar{x}}{(y-\bar{x})^2}-
\frac{2}{y-\bar{x}}-\frac{2}{\bar{x}}+\frac{2}{y \bar{x}^2}+
\frac{2 x}{\bar{y} \bar{x}^2}\right) \ln y \ln\bar{y}
\\
&-\frac{2 \ln  \bar{x} \ln  \bar{y}}{\bar{x} \bar{y}}
+\frac{\ln ^2 x}{\bar{x} \bar{y}}
+\frac{\ln ^2 y}{\bar{x} \bar{y}}
-\frac{\ln ^2 \bar{x}}{\bar{x} \bar{y}}
-\frac{2 \ln ^2 \bar{y}}{\bar{x} \bar{y}}
\\
&+\left(-\frac{4-3 x}{\bar{x}^2 \bar{y}}-\frac{3}{\bar{x}}\right) \ln  x
+\left(\frac{2 (3 x-2)}{\bar{x}^2 \bar{y}}+
 \frac{2 \bar{x}}{(y-\bar{x})^2}+
 \frac{3}{y-\bar{x}}+\frac{3}{\bar{x}}\right) \ln  y
\\
&+\left(\frac{-9 x-1}{x \bar{x} \bar{y}}+\frac{2\bar{x}}{(y-\bar{x})^2}+
  \frac{3}{y-\bar{x}}-\frac{1-3 x}{x^2 y \bar{x}}\right) \ln  \bar{x}
\\
&+\left(-\frac{1}{\bar{x} \bar{y}}-\frac{4}{\bar{x}^2 y}\right) \ln\bar{y}
 +\left(\frac{4}{x \bar{x} \bar{y}}+\frac{4}{x \bar{x}^2 y}\right) \ln (1-x y)
\\
&+\left(-\frac{3 x-1}{x^2 y \bar{x}}+
  \frac{-3 x^2-2 x-1}{x^2 (y-\bar{x})}-
  \frac{3}{\bar{x}}-\frac{2}{\bar{x}^2 \bar{y}}+
  \frac{2 \left(x^2-1\right)}{x (y-\bar{x})^2}\right) \ln (1-x \bar{y})
\\
&+\frac{\pi ^2 \bar{x}^2}{3 (y-\bar{x})^3}+
  \frac{2 \pi ^2 \bar{x}}{3 (y-\bar{x})^2}+
  \frac{\pi ^2}{3 (y-\bar{x})}+\frac{\pi ^2}{3 \bar{x}}-
  \frac{2 \left(2 \pi ^2 x+63 x-63\right)}{3 \bar{y} \bar{x}^2}
\bigg)
\end{split}
\nonumber\\
&&\begin{split}
-\frac{1}{2}C_G\bigg(
&\frac{80 \ln  \frac{\mu }{m_b}}{3 \bar{x} \bar{y}}
+\left(-\frac{2 \ln  x}{\bar{x}^2 \bar{y}}-\frac{22}{3 \bar{x}\bar{y}}\right) 
 \ln  \xi 
\\
&+\left(-\frac{2 x}{(y-x) \bar{x}}+\frac{2 \bar{x}}{(y-\bar{x})^2}+\frac{4}{y-\bar{x}}+\frac{2 (5 x-2)}{\bar{x}^2 \bar{y}}-\frac{2}{y \bar{x}^2}\right) \li x
\\
&+\left(-\frac{2 (x-3)}{\bar{x}^2 \bar{y}}+\frac{2 \bar{x}}{(y-\bar{x})^2}+\frac{4}{y-\bar{x}}+\frac{2 x}{(y-x) \bar{x}}-\frac{4}{\bar{x}}+\frac{2}{y \bar{x}^2}\right) \li y
\\
&+\left(\frac{2 (x-2)}{\bar{x}^2 \bar{y}}+\frac{2}{\bar{x}}\right) \li(x y)
+\left(\frac{2 \bar{x}}{(y-\bar{x})^2}+\frac{4}{y-\bar{x}}+
 \frac{2}{\bar{x}}\right) \li\left(-\frac{x y}{\bar{x}}\right)
\\
&+\left(\frac{2 x}{(y-x) \bar{x}}+\frac{2}{\bar{x}}\right) 
  \li\left(-\frac{y \bar{x}}{\bar{y}}\right)
+\left(-\frac{2}{\bar{x}}+\frac{2}{\bar{x}^2 \bar{y}}+
 \frac{2}{\bar{x}^2 y}\right) \li(x \bar{y})
\\
&+\left(-\frac{2 x}{(y-x) \bar{x}}-\frac{2}{\bar{x}}\right) 
  \li\left(-\frac{x \bar{y}}{\bar{x}}\right)
\\
&+\left(-\frac{2 \bar{x}}{(y-\bar{x})^2}-\frac{4}{y-\bar{x}}-
  \frac{2}{\bar{x}}\right) \li\left(-\frac{\bar{x} \bar{y}}{y}\right)
\\
&-\frac{2 \ln  x \ln  y}{\bar{x} \bar{y}}
+\frac{2 (3 x-2) \ln  x \ln  \bar{x}}{\bar{x}^2 \bar{y}}
+\left(\frac{2}{\bar{x} \bar{y}}-\frac{2}{\bar{x}}\right)\ln y \ln\bar{x}
\\
&+\frac{2 \ln  x \ln  \bar{y}}{\bar{x}^2 \bar{y}}
+\left(\frac{2 \bar{x}}{(y-\bar{x})^2}+\frac{4}{y-\bar{x}}-
  \frac{2}{\bar{x}}+\frac{2}{y \bar{x}^2}+
  \frac{2}{\bar{y} \bar{x}^2}\right) \ln  y \ln  \bar{y}
\\
&+\frac{2 \ln  \bar{x} \ln  \bar{y}}{\bar{x}}
+\frac{\ln ^2 x}{\bar{x} \bar{y}}
+\frac{\ln ^2 y}{\bar{x}}
-\frac{\ln ^2 \bar{x}}{\bar{x} \bar{y}}
-\frac{\ln ^2 \bar{y}}{\bar{x}}
\\
&-\frac{(4-3 x) \ln  x}{\bar{x}^2 \bar{y}}
+\left(-\frac{3-5 x}{\bar{x}^2 \bar{y}}-\frac{2}{y-\bar{x}}\right) \ln  y
\\
&+\left(\frac{2}{x \bar{x} \bar{y}}+\frac{2}{x \bar{x}^2 y}\right) \ln (1-x y)
+\left(\frac{2}{x y \bar{x}}-\frac{31}{3 \bar{x} \bar{y}}-
 \frac{2}{y-\bar{x}}\right) \ln\bar{x}
\\
&-\frac{2 \ln\bar{y}}{y \bar{x}^2}
+\left(\frac{2 (x+1)}{x (y-\bar{x})}-\frac{2}{x y \bar{x}}-
 \frac{2}{\bar{x}^2 \bar{y}}\right) \ln (1-x \bar{y})
\\
&-\frac{2 \left(3 \pi ^2 x+166 x+3 \pi ^2-166\right)}
 {9 \bar{x}^2 \bar{y}}-
 \frac{\pi ^2 \bar{x}}{3 (y-\bar{x})^2}-\frac{2 \pi ^2}{3(y-\bar{x})}+
 \frac{\pi ^2}{3 \bar{x}}
\bigg)\Bigg]
\nonumber\\
\end{split}

%% file: result/T1im.tex
\text{Im}T^{\text{II}(2)}_1&=&-\frac{2\pi\alpha_s^2C_F}{4N_c^2m_B^2\xi}\times
\label{T1im}\\
&&\begin{split}
\Bigg[C_F\bigg(
&-\frac{x \ln  x}{\bar{x}^2 \bar{y}}
+\left(\frac{\bar{x}^2}{(y-\bar{x})^3}+
 \frac{2 \bar{x}}{(y-\bar{x})^2}+\frac{1}{y-\bar{x}}+
 \frac{1}{\bar{y} \bar{x}}\right) \ln  y
\\
&+\left(-\frac{\bar{x}^2}{(y-\bar{x})^3}-
  \frac{2 \bar{x}}{(y-\bar{x})^2}-
  \frac{1}{y-\bar{x}}-\frac{x}{(y-x) \bar{x}}-
  \frac{1}{\bar{y} \bar{x}}\right) \ln  \bar{x}
\\
&+\left(\frac{x}{(y-x) \bar{x}}+\frac{1}{\bar{x} \bar{y}}\right) \ln  \bar{y}
-\frac{\bar{x}}{(y-\bar{x})^2}-
 \frac{3}{2 (y-\bar{x})}+\frac{2}{\bar{y} \bar{x}}
\bigg)
\end{split}
\nonumber\\
&&\begin{split}
-\frac{1}{2}C_G\bigg(
&-\frac{x \ln  x}{\bar{x}^2 \bar{y}}
+\left(-\frac{\bar{x}}{(y-\bar{x})^2}-\frac{2}{y-\bar{x}}\right) \ln  y
\\
&+\left(-\frac{x}{(y-x) \bar{x}}+
  \frac{\bar{x}}{(y-\bar{x})^2}+
  \frac{2}{y-\bar{x}}-\frac{1}{\bar{x} \bar{y}}\right) \ln  \bar{x}
\\
&+\frac{x \ln  \bar{y}}{(y-x) \bar{x}}
+\frac{3}{2 \bar{x} \bar{y}}+\frac{1}{y-\bar{x}}
\bigg)\Bigg]
\\
\end{split}\nonumber

%% file: result/T2re.tex
\text{Re}T^\text{II}_2&=&-\frac{\alpha_s^2C_F C_N}{4N_c^2m_B^2\xi}\times\label{T2re}\\
&&\begin{split}
\Bigg[
&\frac{12 \ln  \frac{\mu }{m_b}}{\bar{x} \bar{y}}
+\left(-\frac{2 \bar{x}^2}{(y-\bar{x})^3}-\frac{4 \bar{x}}{(y-\bar{x})^2}-\frac{2}{y-\bar{x}}-\frac{2 x}{(y-x) \bar{x}}-\frac{2}{\bar{y} \bar{x}}\right) \li x
\\
&+\frac{2 x \li y}{(y-x) \bar{x}}
+\left(-\frac{2 \bar{x}^2}{(y-\bar{x})^3}-\frac{4 \bar{x}}{(y-\bar{x})^2}-\frac{2}{y-\bar{x}}\right) \li\left(-\frac{x y}{\bar{x}}\right)
\\
&+\left(\frac{2 x}{(y-x) \bar{x}}+\frac{2}{\bar{x} \bar{y}}\right) \li\left(-\frac{y \bar{x}}{\bar{y}}\right)
\\
&+\left(\frac{2 \bar{x}^2}{(y-\bar{x})^3}+\frac{4 \bar{x}}{(y-\bar{x})^2}+\frac{2}{y-\bar{x}}+\frac{2}{\bar{y} \bar{x}}\right) \li \bar{y}
\\
&+\left(-\frac{2 x}{(y-x) \bar{x}}-\frac{2}{\bar{x} \bar{y}}\right) \li\left(-\frac{x \bar{y}}{\bar{x}}\right)
\\
&+\left(\frac{2 \bar{x}^2}{(y-\bar{x})^3}+\frac{4 \bar{x}}{(y-\bar{x})^2}+\frac{2}{y-\bar{x}}\right) \li\left(-\frac{\bar{x} \bar{y}}{y}\right)
\\
&-\frac{2 \ln  x \ln  y}{\bar{x} \bar{y}}
+\frac{2 \ln  x \ln  \bar{y}}{\bar{x} \bar{y}}
+\frac{\ln ^2 y}{\bar{x} \bar{y}}
-\frac{\ln ^2 \bar{y}}{\bar{x} \bar{y}}
-\frac{(2-3 x) \ln  x}{\bar{x}^2 \bar{y}}
\\
&+\left(\frac{x-2}{\bar{x}^2 \bar{y}^2}+\frac{2 \bar{x}}{(y-\bar{x})^2}+\frac{3}{y-\bar{x}}+\frac{x}{\bar{x}^2 \bar{y}}\right) \ln  y
\\
&+\left(\frac{2}{x \bar{x} \bar{y}}+\frac{2}{x \bar{x}^2 y}\right) \ln (1-x y)
+\left(\frac{2 \bar{x}}{(y-\bar{x})^2}+\frac{3}{y-\bar{x}}\right) \ln  \bar{x}
\\
&+\left(\frac{-3 x^2-2 x-1}{x^2 (y-\bar{x})}-\frac{1}{x^2 \bar{x}^2 \bar{y}}+\frac{2 \left(x^2-1\right)}{x (y-\bar{x})^2}+\frac{1}{x \bar{x}^2 \bar{y}^2}\right) \ln (1-x \bar{y})
\\
&+\left(-\frac{3}{\bar{x} \bar{y}}-\frac{2}{\bar{x}^2 y}\right) \ln  \bar{y}
+\frac{16}{\bar{x} \bar{y}}
\Bigg]
\end{split}\nonumber

%% file: result/T2im.tex
\text{Im}T^\text{II}_2&=&-\frac{2\pi\alpha_s^2C_F C_N}{4N_c^2m_B^2\xi}
\times\label{T2im}\\
&&\begin{split}
\Bigg[
&\left(\frac{\bar{x}^2}{(y-\bar{x})^3}+\frac{2 \bar{x}}{(y-\bar{x})^2}+\frac{1}{y-\bar{x}}+\frac{1}{\bar{y} \bar{x}}\right) \ln  y
\\
&+\left(-\frac{\bar{x}^2}{(y-\bar{x})^3}-\frac{2 \bar{x}}{(y-\bar{x})^2}-\frac{1}{y-\bar{x}}-\frac{x}{(y-x) \bar{x}}-\frac{1}{\bar{y} \bar{x}}\right) \ln  \bar{x}
\\
&+\frac{x \ln  \bar{y}}{(y-x) \bar{x}}
-\frac{\bar{x}}{(y-\bar{x})^2}-\frac{3}{2 (y-\bar{x})}+\frac{3}{2 \bar{y} \bar{x}}
\nonumber
\Bigg]
\end{split}

%% file: result/convint.tex
By looking at the hard scattering kernels of (\ref{T1re})-(\ref{T2im}) it
is not obvious that there remain no singularities in the convolution
integrals  over
wave functions (\ref{LO4}). It is however possible to perform the
integration analytically, which proves the factorizabilty.

Regarding the $B$-meson wave function we will obtain the result in terms
of the quantities $\lambda_B$ and $\lambda_n$, which are defined in
(\ref{lambdaB}) and (\ref{lambdan}). 
The $\pi$-meson wave function is given in terms of Gegenbauer
polynomials:
\begin{equation}
\phi_\pi(x)=6x\bar{x}\left[1+\sum_{n=1}^\infty
  a_n^\pi C_n^{(3/2)}(2x-1)\right].
\label{gegenbauer}
\end{equation}
Due to the symmetry properties of the pion the first non vanishing
moment is $a_2^\pi$. We neglect $a_n^\pi$ for $n>2$ and 
using (\ref{LO4}) we get for the NLO of $A_\text{spect}$:
\begin{equation}
\begin{split}
&A_\text{spect. 1}^{(2)}=\alpha_s^2\frac{if_\pi^2f_B}{4N_c^2}C_F
\frac{m_B}{\lambda_B}\times
\\
&\begin{split}
\quad\Bigg[&
C_N\left(120\ln\frac{\mu}{m_b}-48\lambda_1+152\right)
\\
&\begin{split}
+C_F\bigg(&(162+36\lambda_1)\ln\frac{\mu}{m_b}-9\lambda_2+
(-54+6\pi^2)\lambda_1+\frac{1566}{5}-\frac{1008}{5}\zeta(3)+27\pi^2
\\
&+i\left(-9\pi+\frac{18}{5}\pi^3\right)
\bigg)
\end{split}
\\
&\begin{split}
-\frac{1}{2}C_G\bigg(&240\ln\frac{\mu}{m_b}+(-102+6\pi^2)\lambda_1+
 \frac{2101}{5}-
 \frac{1008}{5}\zeta(3)+18\pi^2
\\
&+i\left(9\pi+\frac{18}{5}\pi^3\right)
\bigg)
\end{split}
\\
&\begin{split}
+a_2^\pi\bigg\{&
C_N\left(240\ln\frac{\mu}{m_b}-96\lambda_1+404\right)
\\
&\begin{split}
+C_F\bigg(&(174+72\lambda_1)\ln\frac{\mu}{m_b}-18\lambda_2+
\left(-\frac{741}{2}+42\pi^2\right)\lambda_1-\frac{14809}{35}
\\
&-\frac{45072}{35}\zeta(3)+204\pi^2
+i\left(-338\pi+\frac{1362}{35}\pi^3\right)
\bigg)
\end{split}
\\
&\begin{split}
-\frac{1}{2}C_G\bigg(&480\ln\frac{\mu}{m_b}+(-504+42\pi^2)\lambda_1+
\frac{22299}{35}-\frac{43992}{35}\zeta(3)+161\pi^2
\\
&+i\left(-292\pi+\frac{1482}{35}\pi^3\right)\bigg)
\bigg\}
\Bigg]
\end{split}
\end{split}
\end{split}
\end{split}
\label{aspect12}
\end{equation}
and
\begin{equation}
\begin{split}
&A_\text{spect. 2}^{(2)}=\alpha_s^2\frac{if_\pi^2f_B}{4N_c^2}C_FC_N
\frac{m_B}{\lambda_B}\times
\\
&\begin{split}
\quad\Bigg[&
108\ln\frac{\mu}{m_b}+\frac{1467}{10}+\frac{252}{5}\zeta(3)-6\pi^2
+i\left(54\pi-\frac{12}{5}\pi^3\right)
\\
&+a_2^\pi\bigg(216\ln\frac{\mu}{m_b}+\frac{40281}{140}+
\frac{29268}{35}\zeta(3)-112\pi^2+i\left(118\pi-\frac{108}{35}\pi^3\right)
\bigg)\Bigg].
\end{split}
\end{split}
\label{aspect22}
\end{equation}
The finiteness of the above equations proves factorization of the hard
spectator interactions at NLO.

Including the contributions of 
$A_\text{spect. 1}^{(2)}$ and $A_\text{spect. 2}^{(2)}$ 
the quantities
$a_{1,\text{II}}$ and $a_{2,\text{II}}$ defined in (\ref{w13}) and
(\ref{w15}) are
\begin{eqnarray}
a_{1,\text{II}} &=& 
\frac{i}{f_\pi f^{B\pi}_+m_B^2}(C_2A_\text{spect. 1}^{(1)}+
C_2A_\text{spect. 1}^{(2)}
+C_1A_\text{spect. 2}^{(2)})
\nonumber\\
a_{2,\text{II}} &=& 
\frac{i}{f_\pi f^{B\pi}_+ m_B^2}(
C_1A_\text{spect. 1}^{(1)}+
C_1A_\text{spect. 1}^{(2)}
+C_2A_\text{spect. 2}^{(2)}),
\label{aspect3}
\end{eqnarray}
where
\begin{equation}
A_\text{spect. 1}^{(1)}=\frac{-i C_F\pi\alpha_s}{N_c^2}
\frac{f_Bf_\pi^2m_B}{\lambda_B}9(1+a_2^\pi)^2.
\end{equation}

%% file: result/num.tex
\subsection{Input parameters}
%
%
\begin{table}[t]
\begin{tabular*}{\textwidth}{c}
\hline
CKM-parameters\\
\begin{tabular*}{\textwidth}{c@{\extracolsep\fill}cccc}
  $V_\text{ud}$ \cite{Charles:2004jd}& $V_\text{cd}$ & 
  $V_\text{cb}$ \cite{Charles:2004jd}&
  $|V_\text{ub}/V_\text{cb}|$ \cite{Charles:2004jd}& $\gamma$\\
  0.974 & $-0.23$ & 0.041 & 0.09$\pm$0.025 & (70$\pm$20)deg\\
\end{tabular*}\\ \\
Parameters of the $B$-meson\\
\begin{tabular*}{\textwidth}{c@{\extracolsep\fill}cccccc}
  $m_B$ & $f_B$ \cite{Jamin:2001fw}  & $\frac{f_B}{f_+^{B\pi}\lambda_B}$
  \cite{Khodjamirian:2006st} 
  & $\lambda_1$ \cite{Beneke:2005vv} & $\lambda_2$ \cite{Beneke:2005vv}
  & $\tau_{B^\pm}$ & $\tau_{B^0}$ \\
  5.28GeV & \hspace{1ex}(210$\pm$19)MeV & $1.56\pm0.17$ 
  & $-3.2\pm1$ & $11\pm4$ 
  & 1.67 ps & 1.54 ps 
\end{tabular*}\\ \\
Parameters of the $\pi$-meson\\
\begin{tabular*}{\textwidth}{c@{\extracolsep\fill}cccc}
  $f_+^{B\pi}$ \cite{Abada:2000ty,Dalgic:2006dt,Khodjamirian:2000ds}& 
  $f_\pi$ & $m_\pi$ & 
  $a_1^\pi$ & $a_2^\pi$ \cite{Gockeler:2005jz,Braun:2006dg}\\
  $0.28\pm0.05$ & 131MeV & 130MeV &  0 & $0.3\pm0.15$
\end{tabular*}\\ \\ 
Quark and W-boson masses\\
\begin{tabular*}{\textwidth}{c@{\extracolsep\fill}ccc}
  $m_b(m_b)$ & $m_c(m_b)$ & $m_t(m_t)$ \cite{Beneke:2001ev} & $M_W$\\
  4.2 GeV & (1.3$\pm$0.2) GeV & 167 GeV & 80.4 GeV
\end{tabular*}\\ \\
Coupling constants\\
\begin{tabular*}{\textwidth}{c@{\extracolsep\fill}ccc}
  &$\Lambda^{(5)}_{\overline{\text{MS}}}$ & $G_F$ &\\
  &225 MeV & $1.16639\times 10^{-5}\,\text{GeV}^{-2}$&
\end{tabular*}\\
\hline
\end{tabular*}
\caption{Input parameters, which were used in the numerical
  analysis. All parameters given without explicit citation
  can be found in \cite{Yao:2006px}. Unless otherwise stated scale
  dependent quantities are given at $\mu=1\text{GeV}$.} 
\label{params}
\end{table}
For my numerical analysis I use the parameters given in table
\ref{params}. The decay constant $f_B$ and the ratio
$\frac{f_B}{f^{B\pi}_+\lambda_B}$ have been obtained by QCD sum rules
in \cite{Jamin:2001fw} and \cite{Khodjamirian:2006st} respectively. The
logarithmic moments $\lambda_1$ and $\lambda_2$ where calculated in
\cite{Beneke:2005vv} using model light-cone wave functions for the
$B$-meson
\cite{Khodjamirian:2005ea,Braun:2003wx,Grozin:1996pq,Lee:2005gz}.
For the form factor $f_+^{B\pi}$ I use the value from
\cite{Khodjamirian:2000ds}, which has been obtained by QCD sum
rules. This value is consistent with quenched and recent unquenched
lattice calculations \cite{Abada:2000ty,Dalgic:2006dt}. The first
Gegenbauer moment of the pion wave function is zero due to G-parity
while the second moment has been obtained by lattice simulations
\cite{Gockeler:2005jz,Braun:2006dg}.

\subsection{Amplitudes $a_1$ and $a_2$ \label{qcdampl}}
%
\begin{figure}
\begin{center}
\resizebox{0.63\textwidth}{!}{\input{result/numerics/alpha_1_real}}
\resizebox{0.63\textwidth}{!}{\input{result/numerics/alpha_2_real}}
\resizebox{0.63\textwidth}{!}{\input{result/numerics/imag_nlo}}
\end{center}
\caption{Contribution of the hard spectator corrections to $a_1$ and
  $a_2$ as a function of the renormalisation scale $\mu$. The upper
  two figures show the real part, where the LO is given by the dashed 
  line, while the sum of LO and NLO is shown by the thick solid line. 
  The twist-3 corrections are included in the graph given by the thin 
  solid line. The third figure shows the imaginary part, which occurs
  first at $\mathcal{O}(\alpha_s^2)$. So no distinction between LO and
  NLO is made.
}
\label{numpic1}
\end{figure}
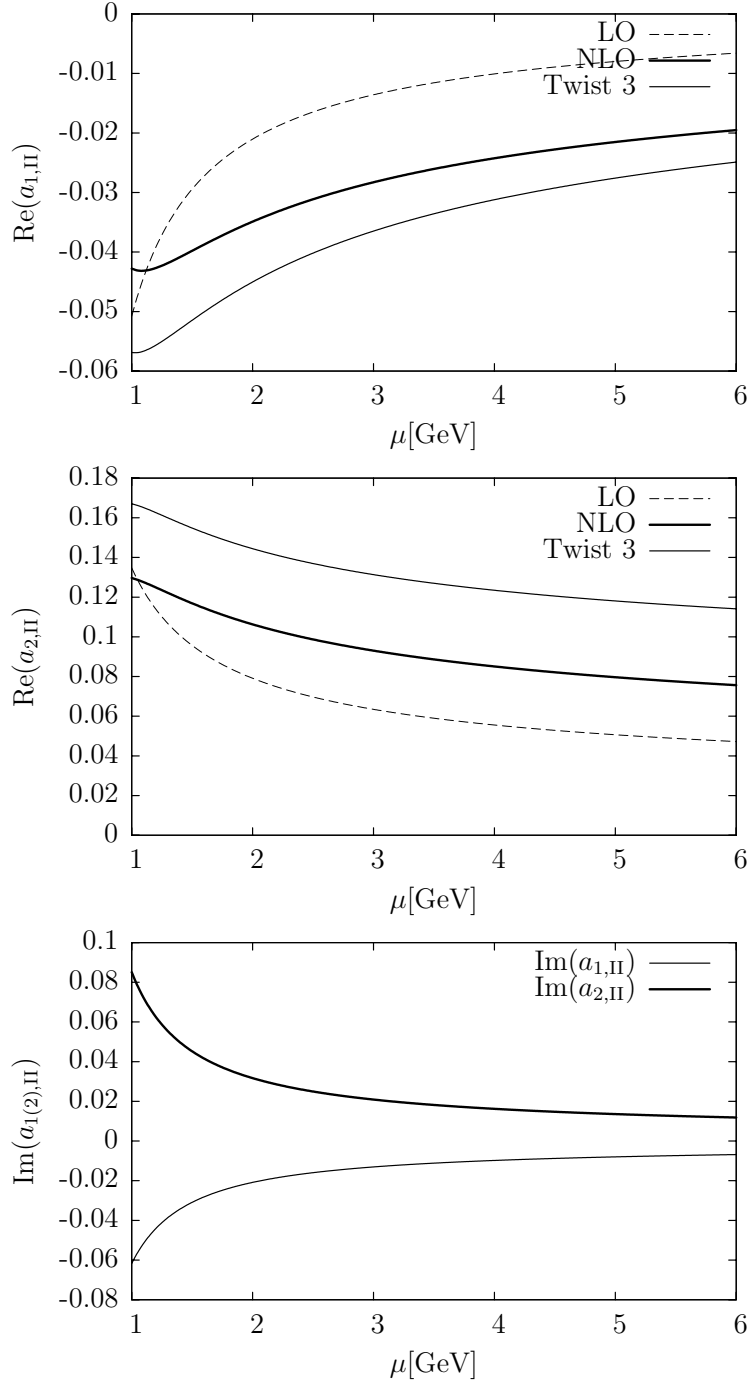
The QCD amplitudes $a_1$ and $a_2$ are defined in \cite{Beneke:2001ev}. 
Their hard
scattering parts $a_{1,\text{II}}$ and $a_{2,\text{II}}$, i.e.\ the
parts of $a_1$ and $a_2$, which contribute to $\mathcal{A}^\text{II}$ 
(see (\ref{w10})), are plotted in fig.~\ref{numpic1} as functions of
the renormalisation scale $\mu$. The strong dependence on $\mu$ of the
real part of LO is reduced at NLO. Taking the twist-3 contributions 
into account does not increase the $\mu$-dependence too much. The
imaginary part, which occurs first at NLO, is strongly dependent on
the renormalisation scale.
An appropriate choice for the scale of the hard scattering amplitude
is the hard collinear scale
\begin{equation}
\mu_\text{hc}=1.5\text{GeV}
\label{num7}
\end{equation} 
In the following numerical calculations we will evaluate
$a_{1,\text{II}}$ and $a_{2,\text{II}}$ at $\mu_\text{hc}$. The vertex 
corrections $\mathcal{A}^\text{I}$ will be evaluated at 
\begin{equation}
\mu_b=4.8\text{GeV}.
\label{num8}
\end{equation}
Using the parameters of table \ref{params} we obtain
\begin{equation}
\begin{split}
a_1 =& 1.015+[0.039+0.018i]_V+[-0.012]_\text{tw3}+
[-0.029]_\text{LO}
\\
&+[-0.010-0.031i]_\text{NLO} 
\\
a_2 =& 0.184+[-0.171-0.080i]_V+[0.038]_\text{tw3}+[0.096]_\text{LO}
\\
&+[0.021+0.045i]_\text{NLO}.
\end{split}
\label{num9}
\end{equation}
These equations are given in a form similar to (61) and
(62) in \cite{Beneke:2005vv}. The first number gives the tree
contribution, the vertex corrections are indicated by the label $V$,
the twist-3 contributions are labelled by $\text{tw3}$. 
The hard scattering part is
separated into LO and NLO. The hadronic input parameters I used are
slightly different from \cite{Beneke:2005vv} and in contrast to
\cite{Beneke:2005vv} I evaluated all quantities, which belong to the hard
scattering amplitude, at the hard collinear scale
$\mu_\text{hc}$. This is why the values I get for $a_1$ and $a_2$ are
different from \cite{Beneke:2005vv}.

The hard scattering amplitudes $a_{1,\text{II}}$ and $a_{2,\text{II}}$
together with their numerical errors read:
\begin{equation}
\begin{split}
a_{1,\text{II}} =\, &
-0.051\pm0.011(\text{param.})^{+0.026}_{-0.005}(\text{scale})
\pm0.012(\text{tw3})\\
&+[-0.031\pm0.008(\text{param.})^{+0.024}_{-0.031}(\text{scale})
\pm0.012(\text{tw3})]i\\
a_{2,\text{II}} =\, &
0.15\pm0.03(\text{param.})^{+0.01}_{-0.04}(\text{scale})
\pm0.04(\text{tw3})\\
&+[0.045\pm0.012(\text{param.})^{+0.040}_{-0.033}(\text{scale})
\pm0.038(\text{tw3})]i.
\end{split}
\label{num10}
\end{equation}
The first error comes from the error of the input parameters in table
\ref{params}. The scale uncertainty is obtained by varying
$\mu_\text{hc}$ between 1GeV and 6GeV. The error labelled by
$\text{tw3}$ gives the error of the twist-3 contribution.
Within the scale uncertainty (\ref{num10}) is compatible
with \cite{Beneke:2005vv}. The result I obtained in QCD comes with
formally large logarithms $\ln\lqcd/m_b$. Without resummation these
logarithms might spoil perturbation theory. However the error coming
from the scale uncertainty in (\ref{num10}) as well as the relative
size of the NLO contributions are small enough for perturbation theory 
to be valid.

\subsection{Branching ratios}
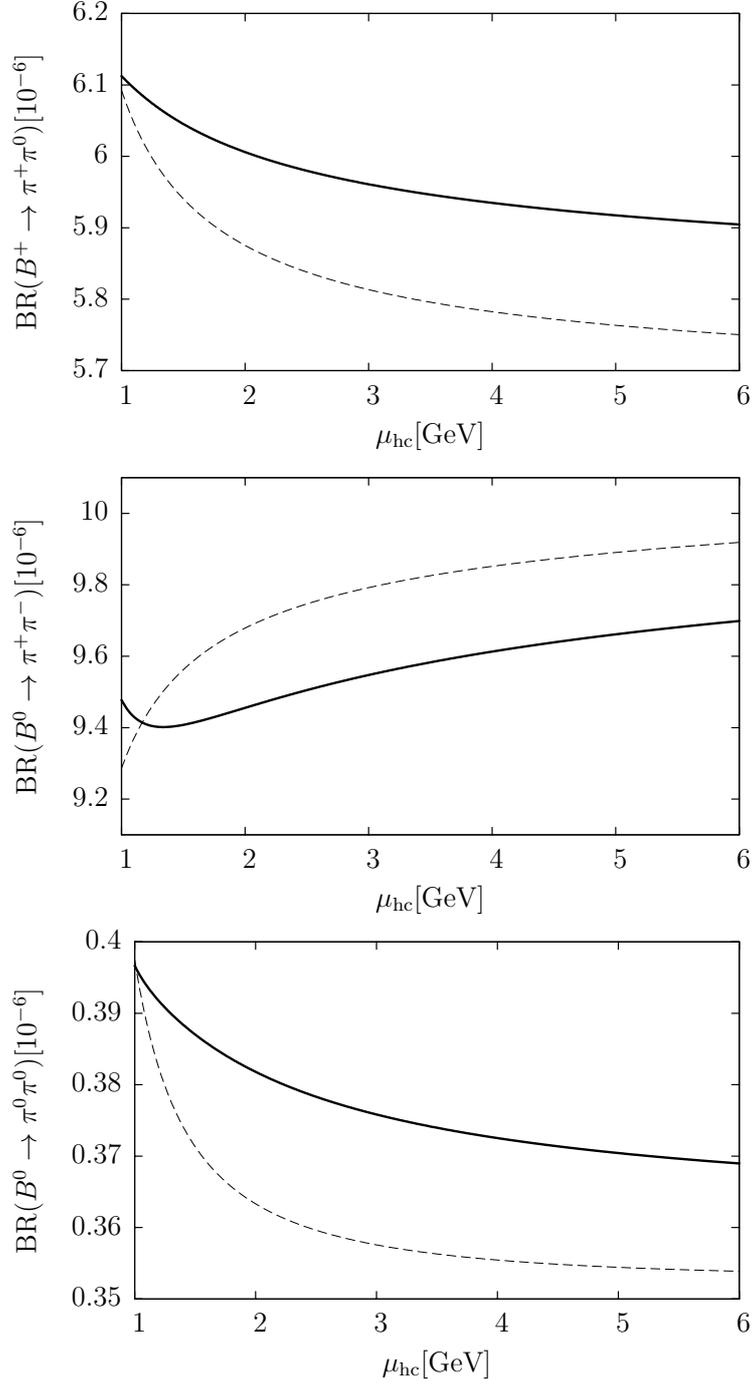
\begin{figure}
\begin{center}
\resizebox{0.63\textwidth}{!}{\input{result/numerics/brp0_scale}}
\resizebox{0.63\textwidth}{!}{\input{result/numerics/brpm_scale}}
\resizebox{0.63\textwidth}{!}{\input{result/numerics/br00_scale}}
\end{center}
\caption{CP-averaged branching ratios as functions of the
  hard collinear scale $\mu_\text{hc}$ in units of $10^{-6}$. 
  In the graph with the dashed line only the leading order of the 
  hard spectator scattering is contained, while in the solid line 
  hard spectator scattering is taken into account up to NLO. 
}
\label{numpic2}
\end{figure}
The dependence of the CP-averaged branching ratios on the hard
collinear scale is shown in fig.~\ref{numpic2}. It is obvious that the
NLO corrections reduce this dependence significantly.

From the parameter set table \ref{params} we obtain the following
CP-averaged branching ratios
\begin{equation}
\begin{split}
10^6\text{BR}(B^+\to\pi^+\pi^0)\;=\;&
6.05^{+2.36}_{-1.98}(\text{had.})^{+2.90}_{-2.33}
(\text{CKM})^{+0.18}_{-0.31}(\text{scale})
\pm0.27(\text{sublead.})\\
10^6\text{BR}(B^0\to\pi^+\pi^-)\;=\;&
9.41^{+3.56}_{-2.99}(\text{had.})^{+4.00}_{-3.46}
(\text{CKM})^{+1.07}_{-3.93}(\text{scale})
^{+1.13}_{-0.70}(\text{sublead.})\\
10^6\text{BR}(B^0\to\pi^0\pi^0)\;=\;&
0.39^{+0.14}_{-0.12}(\text{had.})^{+0.20}_{-0.17}
(\text{CKM})^{+0.17}_{-0.06}(\text{scale})
^{+0.20}_{-0.08}(\text{sublead.}).
\end{split}
\label{num11}
\end{equation}
The origin of the errors are the uncertainties of the hadronic
parameters and the CKM parameters, the scale dependence and the
subleading power contributions, i.e. twist-3 and annihilation
contributions. The error arising from the scale dependence was
estimated by varying $\mu_b$ between 2GeV and 8GeV and $\mu_\text{hc}$
between 1GeV and 6GeV. If we compare (\ref{num11}) to the experimental
values \cite{Barberio:2006bi}:
\begin{eqnarray}
10^6\text{BR}(B^+\to\pi^+\pi^0) &=& 5.5\pm0.6
\nonumber\\
10^6\text{BR}(B^0\to\pi^+\pi^-) &=& 5.0\pm0.4
\nonumber\\
10^6\text{BR}(B^0\to\pi^0\pi^0) &=& 1.45\pm0.29
\label{num12}
\end{eqnarray}
we note that $\text{BR}(B^+\to\pi^+\pi^0)$ is in good agreement with
the data.
For $B^+\to\pi^+\pi^0$ and $B^0\to\pi^+\pi^-$ QCD-factorization is
expected to work well, because at tree level Wilson coefficients occur
in the so called colour allowed combination $C_1+C_2/N_c\sim 1$, while
$B^0\to\pi^0\pi^0$ comes at tree level with $C_2+C_1/N_c\sim 0.2$ such
that subleading power corrections are expected to be more
important. On the other hand there are big uncertainties in the
parameters occurring in the combinations $|V_{ub}|f^{B\pi}_+$,
$\frac{f_B}{f^{B\pi}_+\lambda_B}$ and $a_2^\pi$. In \cite{Beneke:2003zv} 
and \cite{Beneke:2005vv} these parameters were fitted by the 
experimental values (\ref{num12}) of $\text{BR}(B^+\to\pi^+\pi^0)$ 
and $\text{BR}(B^0\to\pi^+\pi^-)$. Setting 
\begin{equation}
a_2^\pi(1\text{GeV})=0.39
\label{num13}
\end{equation} 
leads to 
\begin{eqnarray}
|V_{ub}|f^{B\pi}_+ &\to&  
0.80\left(|V_{ub}|f^{B\pi}_+ \right)_\text{default}
\nonumber\\
\frac{f_B}{f^{B\pi}_+\lambda_B} &\to&
2.89\left(\frac{f_B}{f^{B\pi}_+\lambda_B}\right)_\text{default}.
\label{num14}
\end{eqnarray}
This leads to the following branching ratios:
\begin{eqnarray}
10^6\text{BR}(B^+\to\pi^+\pi^0) &=&
5.5\pm0.2(\text{param.})^{+0.5}_{-0.3}(\text{scale})\pm0.6(\text{sublead.})
\nonumber\\
10^6\text{BR}(B^0\to\pi^+\pi^-) &=&
5.0^{+0.8}_{-0.9}(\text{param.})^{+0.9}_{-0.2}(\text{scale})
^{+0.9}_{-0.6}(\text{sublead.})
\nonumber\\
10^6\text{BR}(B^0\to\pi^0\pi^0) &=&
0.77\pm 0.3(\text{param.})^{+0.2}_{-0.3}(\text{scale})
^{+0.3}_{-0.2}(\text{sublead.}).
\label{num15}
\end{eqnarray}
The uncertainties of the quantities that occurred in (\ref{num13}) 
and (\ref{num14}) have not been considered in the estimation of the
errors in (\ref{num15}). The $B^0\to\pi^0\pi^0$ branching ratio
obtained in (\ref{num15}) is compatible with the value obtained in 
\cite{Beneke:2005vv}. Though it is too low, due to the
theoretical and experimental errors it is compatible with
(\ref{num12}).

%% file: result/numerics/alpha_1_real.tex
\begingroup
  \makeatletter
  \providecommand\color[2][]{%
    \GenericError{(gnuplot) \space\space\space\@spaces}{%
      Package color not loaded in conjunction with
      terminal option `colourtext'%
    }{See the gnuplot documentation for explanation.%
    }{Either use 'blacktext' in gnuplot or load the package
      color.sty in LaTeX.}%
    \renewcommand\color[2][]{}%
  }%
  \providecommand\includegraphics[2][]{%
    \GenericError{(gnuplot) \space\space\space\@spaces}{%
      Package graphicx or graphics not loaded%
    }{See the gnuplot documentation for explanation.%
    }{The gnuplot epslatex terminal needs graphicx.sty or graphics.sty.}%
    \renewcommand\includegraphics[2][]{}%
  }%
  \providecommand\rotatebox[2]{#2}%
  \@ifundefined{ifGPcolor}{%
    \newif\ifGPcolor
    \GPcolorfalse
  }{}%
  \@ifundefined{ifGPblacktext}{%
    \newif\ifGPblacktext
    \GPblacktextfalse
  }{}%
  \let\gplgaddtomacro\g@addto@macro
  \gdef\gplbacktext{}%
  \gdef\gplfronttext{}%
  \makeatother
  \ifGPblacktext
    \def\colorrgb#1{}%
    \def\colorgray#1{}%
  \else
    \ifGPcolor
      \def\colorrgb#1{\color[rgb]{#1}}%
      \def\colorgray#1{\color[gray]{#1}}%
      \expandafter\def\csname LTw\endcsname{\color{white}}%
      \expandafter\def\csname LTb\endcsname{\color{black}}%
      \expandafter\def\csname LTa\endcsname{\color{black}}%
      \expandafter\def\csname LT0\endcsname{\color[rgb]{1,0,0}}%
      \expandafter\def\csname LT1\endcsname{\color[rgb]{0,0,1}}%
      \expandafter\def\csname LT2\endcsname{\color[rgb]{0,1,1}}%
      \expandafter\def\csname LT3\endcsname{\color[rgb]{1,0,1}}%
    \else
      \def\colorrgb#1{\color{black}}%
      \def\colorgray#1{\color[gray]{#1}}%
      \expandafter\def\csname LTw\endcsname{\color{white}}%
      \expandafter\def\csname LTb\endcsname{\color{black}}%
      \expandafter\def\csname LTa\endcsname{\color{black}}%
      \expandafter\def\csname LT0\endcsname{\color{black}}%
      \expandafter\def\csname LT1\endcsname{\color{black}}%
      \expandafter\def\csname LT2\endcsname{\color{black}}%
      \expandafter\def\csname LT3\endcsname{\color{black}}%
      \expandafter\def\csname LT4\endcsname{\color{black}}%
      \expandafter\def\csname LT5\endcsname{\color{black}}%
      \expandafter\def\csname LT6\endcsname{\color{black}}%
      \expandafter\def\csname LT7\endcsname{\color{black}}%
      \expandafter\def\csname LT8\endcsname{\color{black}}%
    \fi
  \fi
  \setlength{\unitlength}{0.0500bp}%
  \begin{picture}(6480.00,3888.00)%
    \gplgaddtomacro\gplbacktext{%
      \csname LTb\endcsname%
      \put(990,660){\makebox(0,0)[r]{\strut{}-0.06}}%
      \put(990,1161){\makebox(0,0)[r]{\strut{}-0.05}}%
      \put(990,1663){\makebox(0,0)[r]{\strut{}-0.04}}%
      \put(990,2164){\makebox(0,0)[r]{\strut{}-0.03}}%
      \put(990,2665){\makebox(0,0)[r]{\strut{}-0.02}}%
      \put(990,3167){\makebox(0,0)[r]{\strut{}-0.01}}%
      \put(990,3668){\makebox(0,0)[r]{\strut{} 0}}%
      \put(1100,440){\makebox(0,0){\strut{} 1}}%
      \put(2110,440){\makebox(0,0){\strut{} 2}}%
      \put(3120,440){\makebox(0,0){\strut{} 3}}%
      \put(4130,440){\makebox(0,0){\strut{} 4}}%
      \put(5140,440){\makebox(0,0){\strut{} 5}}%
      \put(6150,440){\makebox(0,0){\strut{} 6}}%
      \put(220,2164){\rotatebox{90}{\makebox(0,0){\strut{}$\text{Re}(a_{1,\text{II}})$}}}%
      \put(3625,110){\makebox(0,0){\strut{}$\mu[\mbox{GeV}]$}}%
    }%
    \gplgaddtomacro\gplfronttext{%
      \csname LTb\endcsname%
      \put(5317,3495){\makebox(0,0)[r]{\strut{}LO}}%
      \csname LTb\endcsname%
      \put(5317,3275){\makebox(0,0)[r]{\strut{}NLO}}%
      \put(5317,3055){\makebox(0,0)[r]{\strut{}Twist 3}}%
    }%
    \gplbacktext
    \put(0,0){\includegraphics{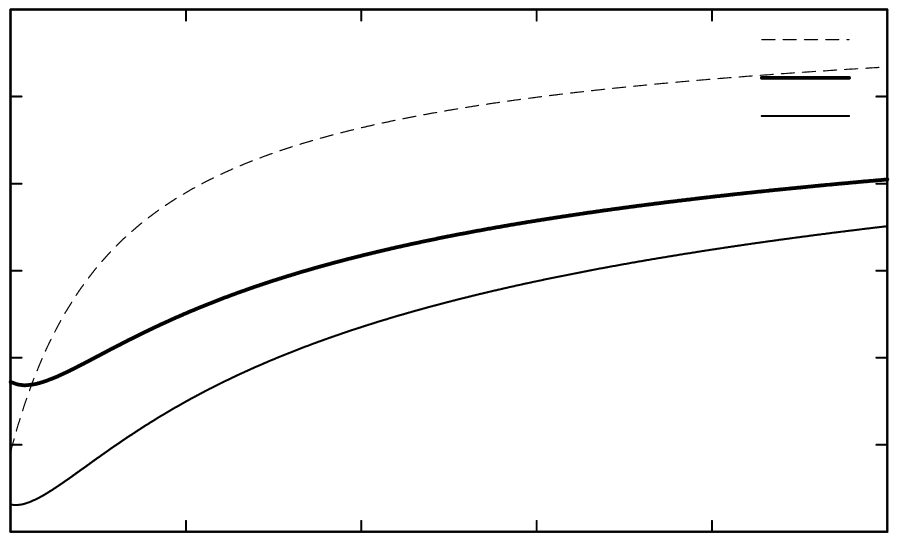}}%
    \gplfronttext
  \end{picture}%
\endgroup

%% file: result/numerics/alpha_2_real.tex
\begingroup
  \makeatletter
  \providecommand\color[2][]{%
    \GenericError{(gnuplot) \space\space\space\@spaces}{%
      Package color not loaded in conjunction with
      terminal option `colourtext'%
    }{See the gnuplot documentation for explanation.%
    }{Either use 'blacktext' in gnuplot or load the package
      color.sty in LaTeX.}%
    \renewcommand\color[2][]{}%
  }%
  \providecommand\includegraphics[2][]{%
    \GenericError{(gnuplot) \space\space\space\@spaces}{%
      Package graphicx or graphics not loaded%
    }{See the gnuplot documentation for explanation.%
    }{The gnuplot epslatex terminal needs graphicx.sty or graphics.sty.}%
    \renewcommand\includegraphics[2][]{}%
  }%
  \providecommand\rotatebox[2]{#2}%
  \@ifundefined{ifGPcolor}{%
    \newif\ifGPcolor
    \GPcolorfalse
  }{}%
  \@ifundefined{ifGPblacktext}{%
    \newif\ifGPblacktext
    \GPblacktextfalse
  }{}%
  \let\gplgaddtomacro\g@addto@macro
  \gdef\gplbacktext{}%
  \gdef\gplfronttext{}%
  \makeatother
  \ifGPblacktext
    \def\colorrgb#1{}%
    \def\colorgray#1{}%
  \else
    \ifGPcolor
      \def\colorrgb#1{\color[rgb]{#1}}%
      \def\colorgray#1{\color[gray]{#1}}%
      \expandafter\def\csname LTw\endcsname{\color{white}}%
      \expandafter\def\csname LTb\endcsname{\color{black}}%
      \expandafter\def\csname LTa\endcsname{\color{black}}%
      \expandafter\def\csname LT0\endcsname{\color[rgb]{1,0,0}}%
      \expandafter\def\csname LT1\endcsname{\color[rgb]{0,0,1}}%
      \expandafter\def\csname LT2\endcsname{\color[rgb]{0,1,1}}%
      \expandafter\def\csname LT3\endcsname{\color[rgb]{1,0,1}}%
    \else
      \def\colorrgb#1{\color{black}}%
      \def\colorgray#1{\color[gray]{#1}}%
      \expandafter\def\csname LTw\endcsname{\color{white}}%
      \expandafter\def\csname LTb\endcsname{\color{black}}%
      \expandafter\def\csname LTa\endcsname{\color{black}}%
      \expandafter\def\csname LT0\endcsname{\color{black}}%
      \expandafter\def\csname LT1\endcsname{\color{black}}%
      \expandafter\def\csname LT2\endcsname{\color{black}}%
      \expandafter\def\csname LT3\endcsname{\color{black}}%
      \expandafter\def\csname LT4\endcsname{\color{black}}%
      \expandafter\def\csname LT5\endcsname{\color{black}}%
      \expandafter\def\csname LT6\endcsname{\color{black}}%
      \expandafter\def\csname LT7\endcsname{\color{black}}%
      \expandafter\def\csname LT8\endcsname{\color{black}}%
    \fi
  \fi
  \setlength{\unitlength}{0.0500bp}%
  \begin{picture}(6480.00,3888.00)%
    \gplgaddtomacro\gplbacktext{%
      \csname LTb\endcsname%
      \put(990,660){\makebox(0,0)[r]{\strut{} 0}}%
      \put(990,994){\makebox(0,0)[r]{\strut{} 0.02}}%
      \put(990,1328){\makebox(0,0)[r]{\strut{} 0.04}}%
      \put(990,1663){\makebox(0,0)[r]{\strut{} 0.06}}%
      \put(990,1997){\makebox(0,0)[r]{\strut{} 0.08}}%
      \put(990,2331){\makebox(0,0)[r]{\strut{} 0.1}}%
      \put(990,2665){\makebox(0,0)[r]{\strut{} 0.12}}%
      \put(990,3000){\makebox(0,0)[r]{\strut{} 0.14}}%
      \put(990,3334){\makebox(0,0)[r]{\strut{} 0.16}}%
      \put(990,3668){\makebox(0,0)[r]{\strut{} 0.18}}%
      \put(1100,440){\makebox(0,0){\strut{} 1}}%
      \put(2110,440){\makebox(0,0){\strut{} 2}}%
      \put(3120,440){\makebox(0,0){\strut{} 3}}%
      \put(4130,440){\makebox(0,0){\strut{} 4}}%
      \put(5140,440){\makebox(0,0){\strut{} 5}}%
      \put(6150,440){\makebox(0,0){\strut{} 6}}%
      \put(220,2164){\rotatebox{90}{\makebox(0,0){\strut{}$\text{Re}(a_{2,\text{II}})$}}}%
      \put(3625,110){\makebox(0,0){\strut{}$\mu[\mbox{GeV}]$}}%
    }%
    \gplgaddtomacro\gplfronttext{%
      \csname LTb\endcsname%
      \put(5317,3495){\makebox(0,0)[r]{\strut{}LO}}%
      \csname LTb\endcsname%
      \put(5317,3275){\makebox(0,0)[r]{\strut{}NLO}}%
      \put(5317,3055){\makebox(0,0)[r]{\strut{}Twist 3}}%
    }%
    \gplbacktext
    \put(0,0){\includegraphics{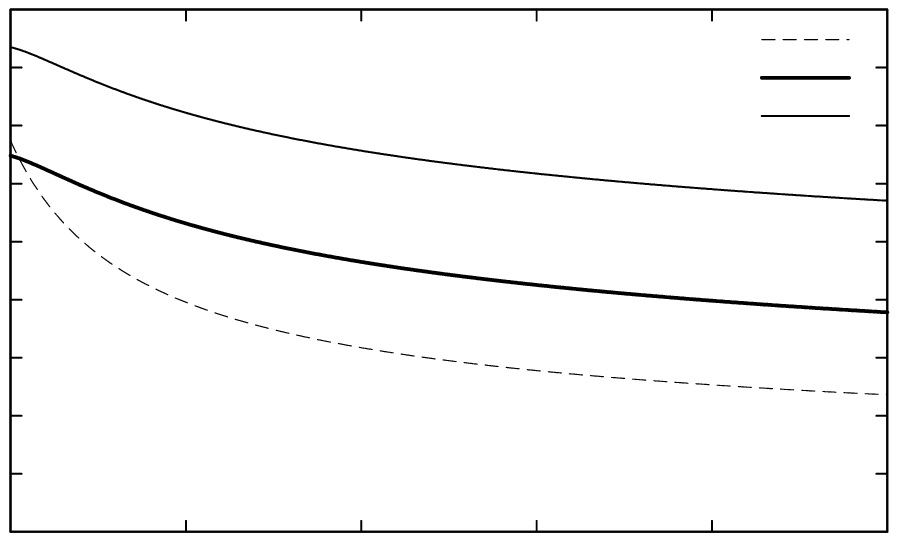}}%
    \gplfronttext
  \end{picture}%
\endgroup

%% file: result/numerics/imag_nlo.tex
\begingroup
  \makeatletter
  \providecommand\color[2][]{%
    \GenericError{(gnuplot) \space\space\space\@spaces}{%
      Package color not loaded in conjunction with
      terminal option `colourtext'%
    }{See the gnuplot documentation for explanation.%
    }{Either use 'blacktext' in gnuplot or load the package
      color.sty in LaTeX.}%
    \renewcommand\color[2][]{}%
  }%
  \providecommand\includegraphics[2][]{%
    \GenericError{(gnuplot) \space\space\space\@spaces}{%
      Package graphicx or graphics not loaded%
    }{See the gnuplot documentation for explanation.%
    }{The gnuplot epslatex terminal needs graphicx.sty or graphics.sty.}%
    \renewcommand\includegraphics[2][]{}%
  }%
  \providecommand\rotatebox[2]{#2}%
  \@ifundefined{ifGPcolor}{%
    \newif\ifGPcolor
    \GPcolorfalse
  }{}%
  \@ifundefined{ifGPblacktext}{%
    \newif\ifGPblacktext
    \GPblacktextfalse
  }{}%
  \let\gplgaddtomacro\g@addto@macro
  \gdef\gplbacktext{}%
  \gdef\gplfronttext{}%
  \makeatother
  \ifGPblacktext
    \def\colorrgb#1{}%
    \def\colorgray#1{}%
  \else
    \ifGPcolor
      \def\colorrgb#1{\color[rgb]{#1}}%
      \def\colorgray#1{\color[gray]{#1}}%
      \expandafter\def\csname LTw\endcsname{\color{white}}%
      \expandafter\def\csname LTb\endcsname{\color{black}}%
      \expandafter\def\csname LTa\endcsname{\color{black}}%
      \expandafter\def\csname LT0\endcsname{\color[rgb]{1,0,0}}%
      \expandafter\def\csname LT1\endcsname{\color[rgb]{0,0,1}}%
      \expandafter\def\csname LT2\endcsname{\color[rgb]{0,1,1}}%
      \expandafter\def\csname LT3\endcsname{\color[rgb]{1,0,1}}%
    \else
      \def\colorrgb#1{\color{black}}%
      \def\colorgray#1{\color[gray]{#1}}%
      \expandafter\def\csname LTw\endcsname{\color{white}}%
      \expandafter\def\csname LTb\endcsname{\color{black}}%
      \expandafter\def\csname LTa\endcsname{\color{black}}%
      \expandafter\def\csname LT0\endcsname{\color{black}}%
      \expandafter\def\csname LT1\endcsname{\color{black}}%
      \expandafter\def\csname LT2\endcsname{\color{black}}%
      \expandafter\def\csname LT3\endcsname{\color{black}}%
      \expandafter\def\csname LT4\endcsname{\color{black}}%
      \expandafter\def\csname LT5\endcsname{\color{black}}%
      \expandafter\def\csname LT6\endcsname{\color{black}}%
      \expandafter\def\csname LT7\endcsname{\color{black}}%
      \expandafter\def\csname LT8\endcsname{\color{black}}%
    \fi
  \fi
  \setlength{\unitlength}{0.0500bp}%
  \begin{picture}(6480.00,3888.00)%
    \gplgaddtomacro\gplbacktext{%
      \csname LTb\endcsname%
      \put(990,660){\makebox(0,0)[r]{\strut{}-0.08}}%
      \put(990,994){\makebox(0,0)[r]{\strut{}-0.06}}%
      \put(990,1328){\makebox(0,0)[r]{\strut{}-0.04}}%
      \put(990,1663){\makebox(0,0)[r]{\strut{}-0.02}}%
      \put(990,1997){\makebox(0,0)[r]{\strut{} 0}}%
      \put(990,2331){\makebox(0,0)[r]{\strut{} 0.02}}%
      \put(990,2665){\makebox(0,0)[r]{\strut{} 0.04}}%
      \put(990,3000){\makebox(0,0)[r]{\strut{} 0.06}}%
      \put(990,3334){\makebox(0,0)[r]{\strut{} 0.08}}%
      \put(990,3668){\makebox(0,0)[r]{\strut{} 0.1}}%
      \put(1100,440){\makebox(0,0){\strut{} 1}}%
      \put(2110,440){\makebox(0,0){\strut{} 2}}%
      \put(3120,440){\makebox(0,0){\strut{} 3}}%
      \put(4130,440){\makebox(0,0){\strut{} 4}}%
      \put(5140,440){\makebox(0,0){\strut{} 5}}%
      \put(6150,440){\makebox(0,0){\strut{} 6}}%
      \put(220,2164){\rotatebox{90}{\makebox(0,0){\strut{}$\text{Im}(a_{1(2),\text{II}})$}}}%
      \put(3625,110){\makebox(0,0){\strut{}$\mu[\mbox{GeV}]$}}%
    }%
    \gplgaddtomacro\gplfronttext{%
      \put(5317,3495){\makebox(0,0)[r]{\strut{}$\text{Im}(a_{1,\text{II}})$}}%
      \put(5317,3275){\makebox(0,0)[r]{\strut{}$\text{Im}(a_{2,\text{II}})$}}%
    }%
    \gplbacktext
    \put(0,0){\includegraphics{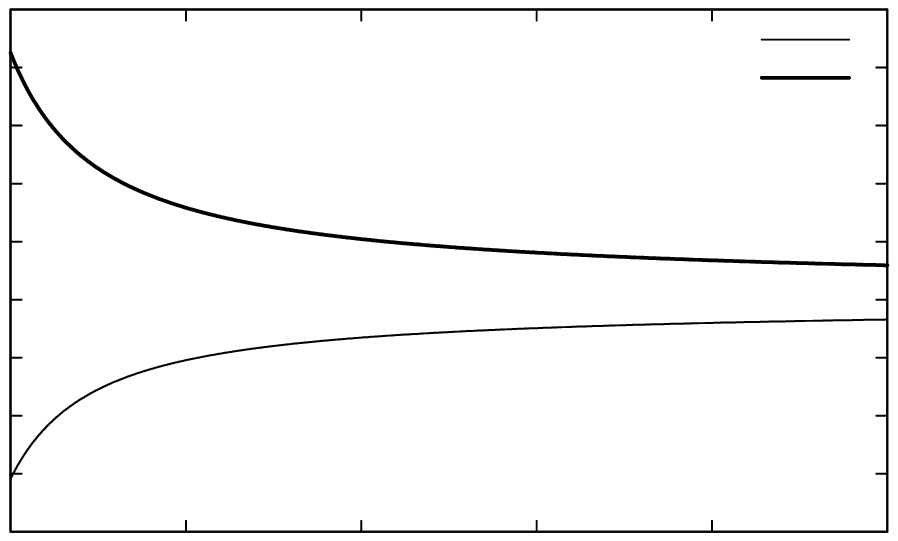}}%
    \gplfronttext
  \end{picture}%
\endgroup

%% file: result/numerics/brp0_scale.tex
\begingroup
  \makeatletter
  \providecommand\color[2][]{%
    \GenericError{(gnuplot) \space\space\space\@spaces}{%
      Package color not loaded in conjunction with
      terminal option `colourtext'%
    }{See the gnuplot documentation for explanation.%
    }{Either use 'blacktext' in gnuplot or load the package
      color.sty in LaTeX.}%
    \renewcommand\color[2][]{}%
  }%
  \providecommand\includegraphics[2][]{%
    \GenericError{(gnuplot) \space\space\space\@spaces}{%
      Package graphicx or graphics not loaded%
    }{See the gnuplot documentation for explanation.%
    }{The gnuplot epslatex terminal needs graphicx.sty or graphics.sty.}%
    \renewcommand\includegraphics[2][]{}%
  }%
  \providecommand\rotatebox[2]{#2}%
  \@ifundefined{ifGPcolor}{%
    \newif\ifGPcolor
    \GPcolorfalse
  }{}%
  \@ifundefined{ifGPblacktext}{%
    \newif\ifGPblacktext
    \GPblacktextfalse
  }{}%
  \let\gplgaddtomacro\g@addto@macro
  \gdef\gplbacktext{}%
  \gdef\gplfronttext{}%
  \makeatother
  \ifGPblacktext
    \def\colorrgb#1{}%
    \def\colorgray#1{}%
  \else
    \ifGPcolor
      \def\colorrgb#1{\color[rgb]{#1}}%
      \def\colorgray#1{\color[gray]{#1}}%
      \expandafter\def\csname LTw\endcsname{\color{white}}%
      \expandafter\def\csname LTb\endcsname{\color{black}}%
      \expandafter\def\csname LTa\endcsname{\color{black}}%
      \expandafter\def\csname LT0\endcsname{\color[rgb]{1,0,0}}%
      \expandafter\def\csname LT1\endcsname{\color[rgb]{0,0,1}}%
      \expandafter\def\csname LT2\endcsname{\color[rgb]{0,1,1}}%
      \expandafter\def\csname LT3\endcsname{\color[rgb]{1,0,1}}%
    \else
      \def\colorrgb#1{\color{black}}%
      \def\colorgray#1{\color[gray]{#1}}%
      \expandafter\def\csname LTw\endcsname{\color{white}}%
      \expandafter\def\csname LTb\endcsname{\color{black}}%
      \expandafter\def\csname LTa\endcsname{\color{black}}%
      \expandafter\def\csname LT0\endcsname{\color{black}}%
      \expandafter\def\csname LT1\endcsname{\color{black}}%
      \expandafter\def\csname LT2\endcsname{\color{black}}%
      \expandafter\def\csname LT3\endcsname{\color{black}}%
      \expandafter\def\csname LT4\endcsname{\color{black}}%
      \expandafter\def\csname LT5\endcsname{\color{black}}%
      \expandafter\def\csname LT6\endcsname{\color{black}}%
      \expandafter\def\csname LT7\endcsname{\color{black}}%
      \expandafter\def\csname LT8\endcsname{\color{black}}%
    \fi
  \fi
  \setlength{\unitlength}{0.0500bp}%
  \begin{picture}(6480.00,3888.00)%
    \gplgaddtomacro\gplbacktext{%
      \csname LTb\endcsname%
      \put(880,660){\makebox(0,0)[r]{\strut{} 5.7}}%
      \put(880,1262){\makebox(0,0)[r]{\strut{} 5.8}}%
      \put(880,1863){\makebox(0,0)[r]{\strut{} 5.9}}%
      \put(880,2465){\makebox(0,0)[r]{\strut{} 6}}%
      \put(880,3066){\makebox(0,0)[r]{\strut{} 6.1}}%
      \put(880,3668){\makebox(0,0)[r]{\strut{} 6.2}}%
      \put(990,440){\makebox(0,0){\strut{} 1}}%
      \put(2022,440){\makebox(0,0){\strut{} 2}}%
      \put(3054,440){\makebox(0,0){\strut{} 3}}%
      \put(4086,440){\makebox(0,0){\strut{} 4}}%
      \put(5118,440){\makebox(0,0){\strut{} 5}}%
      \put(6150,440){\makebox(0,0){\strut{} 6}}%
      \put(220,2164){\rotatebox{90}{\makebox(0,0){\strut{}$\text{BR}(B^+\to\pi^+\pi^0)[10^{-6}]$}}}%
      \put(3570,110){\makebox(0,0){\strut{}$\mu_\text{hc}[\mbox{GeV}]$}}%
    }%
    \gplgaddtomacro\gplfronttext{%
    }%
    \gplbacktext
    \put(0,0){\includegraphics{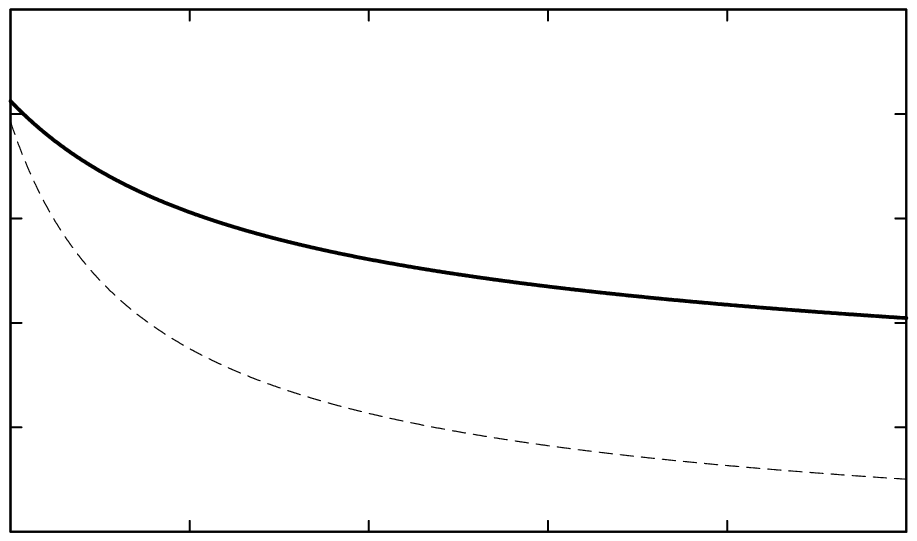}}%
    \gplfronttext
  \end{picture}%
\endgroup

%% file: result/numerics/brpm_scale.tex
\begingroup
  \makeatletter
  \providecommand\color[2][]{%
    \GenericError{(gnuplot) \space\space\space\@spaces}{%
      Package color not loaded in conjunction with
      terminal option `colourtext'%
    }{See the gnuplot documentation for explanation.%
    }{Either use 'blacktext' in gnuplot or load the package
      color.sty in LaTeX.}%
    \renewcommand\color[2][]{}%
  }%
  \providecommand\includegraphics[2][]{%
    \GenericError{(gnuplot) \space\space\space\@spaces}{%
      Package graphicx or graphics not loaded%
    }{See the gnuplot documentation for explanation.%
    }{The gnuplot epslatex terminal needs graphicx.sty or graphics.sty.}%
    \renewcommand\includegraphics[2][]{}%
  }%
  \providecommand\rotatebox[2]{#2}%
  \@ifundefined{ifGPcolor}{%
    \newif\ifGPcolor
    \GPcolorfalse
  }{}%
  \@ifundefined{ifGPblacktext}{%
    \newif\ifGPblacktext
    \GPblacktextfalse
  }{}%
  \let\gplgaddtomacro\g@addto@macro
  \gdef\gplbacktext{}%
  \gdef\gplfronttext{}%
  \makeatother
  \ifGPblacktext
    \def\colorrgb#1{}%
    \def\colorgray#1{}%
  \else
    \ifGPcolor
      \def\colorrgb#1{\color[rgb]{#1}}%
      \def\colorgray#1{\color[gray]{#1}}%
      \expandafter\def\csname LTw\endcsname{\color{white}}%
      \expandafter\def\csname LTb\endcsname{\color{black}}%
      \expandafter\def\csname LTa\endcsname{\color{black}}%
      \expandafter\def\csname LT0\endcsname{\color[rgb]{1,0,0}}%
      \expandafter\def\csname LT1\endcsname{\color[rgb]{0,0,1}}%
      \expandafter\def\csname LT2\endcsname{\color[rgb]{0,1,1}}%
      \expandafter\def\csname LT3\endcsname{\color[rgb]{1,0,1}}%
    \else
      \def\colorrgb#1{\color{black}}%
      \def\colorgray#1{\color[gray]{#1}}%
      \expandafter\def\csname LTw\endcsname{\color{white}}%
      \expandafter\def\csname LTb\endcsname{\color{black}}%
      \expandafter\def\csname LTa\endcsname{\color{black}}%
      \expandafter\def\csname LT0\endcsname{\color{black}}%
      \expandafter\def\csname LT1\endcsname{\color{black}}%
      \expandafter\def\csname LT2\endcsname{\color{black}}%
      \expandafter\def\csname LT3\endcsname{\color{black}}%
      \expandafter\def\csname LT4\endcsname{\color{black}}%
      \expandafter\def\csname LT5\endcsname{\color{black}}%
      \expandafter\def\csname LT6\endcsname{\color{black}}%
      \expandafter\def\csname LT7\endcsname{\color{black}}%
      \expandafter\def\csname LT8\endcsname{\color{black}}%
    \fi
  \fi
  \setlength{\unitlength}{0.0500bp}%
  \begin{picture}(6480.00,3888.00)%
    \gplgaddtomacro\gplbacktext{%
      \csname LTb\endcsname%
      \put(880,961){\makebox(0,0)[r]{\strut{} 9.2}}%
      \put(880,1562){\makebox(0,0)[r]{\strut{} 9.4}}%
      \put(880,2164){\makebox(0,0)[r]{\strut{} 9.6}}%
      \put(880,2766){\makebox(0,0)[r]{\strut{} 9.8}}%
      \put(880,3367){\makebox(0,0)[r]{\strut{} 10}}%
      \put(990,440){\makebox(0,0){\strut{} 1}}%
      \put(2022,440){\makebox(0,0){\strut{} 2}}%
      \put(3054,440){\makebox(0,0){\strut{} 3}}%
      \put(4086,440){\makebox(0,0){\strut{} 4}}%
      \put(5118,440){\makebox(0,0){\strut{} 5}}%
      \put(6150,440){\makebox(0,0){\strut{} 6}}%
      \put(220,2164){\rotatebox{90}{\makebox(0,0){\strut{}$\text{BR}(B^0\to\pi^+\pi^-)[10^{-6}]$}}}%
      \put(3570,110){\makebox(0,0){\strut{}$\mu_\text{hc}[\mbox{GeV}]$}}%
    }%
    \gplgaddtomacro\gplfronttext{%
    }%
    \gplbacktext
    \put(0,0){\includegraphics{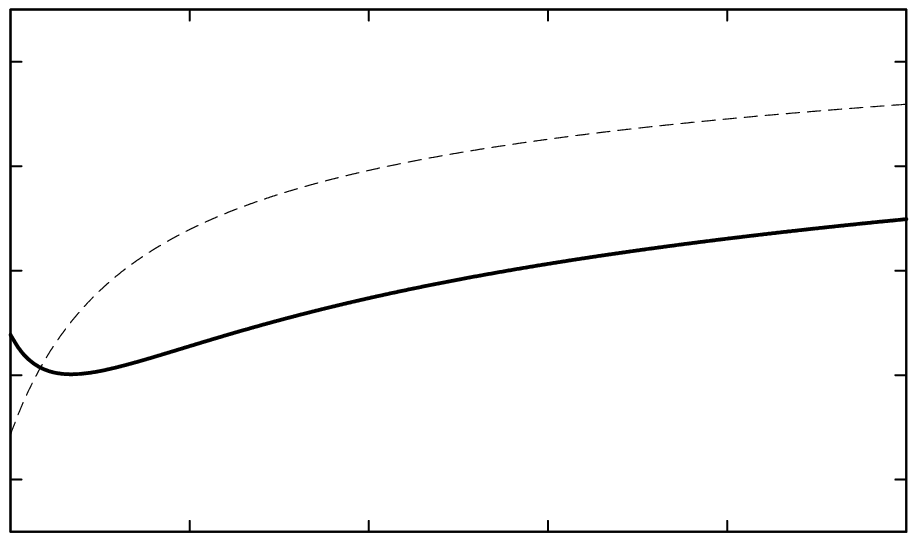}}%
    \gplfronttext
  \end{picture}%
\endgroup

%% file: result/numerics/br00_scale.tex
\begingroup
  \makeatletter
  \providecommand\color[2][]{%
    \GenericError{(gnuplot) \space\space\space\@spaces}{%
      Package color not loaded in conjunction with
      terminal option `colourtext'%
    }{See the gnuplot documentation for explanation.%
    }{Either use 'blacktext' in gnuplot or load the package
      color.sty in LaTeX.}%
    \renewcommand\color[2][]{}%
  }%
  \providecommand\includegraphics[2][]{%
    \GenericError{(gnuplot) \space\space\space\@spaces}{%
      Package graphicx or graphics not loaded%
    }{See the gnuplot documentation for explanation.%
    }{The gnuplot epslatex terminal needs graphicx.sty or graphics.sty.}%
    \renewcommand\includegraphics[2][]{}%
  }%
  \providecommand\rotatebox[2]{#2}%
  \@ifundefined{ifGPcolor}{%
    \newif\ifGPcolor
    \GPcolorfalse
  }{}%
  \@ifundefined{ifGPblacktext}{%
    \newif\ifGPblacktext
    \GPblacktextfalse
  }{}%
  \let\gplgaddtomacro\g@addto@macro
  \gdef\gplbacktext{}%
  \gdef\gplfronttext{}%
  \makeatother
  \ifGPblacktext
    \def\colorrgb#1{}%
    \def\colorgray#1{}%
  \else
    \ifGPcolor
      \def\colorrgb#1{\color[rgb]{#1}}%
      \def\colorgray#1{\color[gray]{#1}}%
      \expandafter\def\csname LTw\endcsname{\color{white}}%
      \expandafter\def\csname LTb\endcsname{\color{black}}%
      \expandafter\def\csname LTa\endcsname{\color{black}}%
      \expandafter\def\csname LT0\endcsname{\color[rgb]{1,0,0}}%
      \expandafter\def\csname LT1\endcsname{\color[rgb]{0,0,1}}%
      \expandafter\def\csname LT2\endcsname{\color[rgb]{0,1,1}}%
      \expandafter\def\csname LT3\endcsname{\color[rgb]{1,0,1}}%
    \else
      \def\colorrgb#1{\color{black}}%
      \def\colorgray#1{\color[gray]{#1}}%
      \expandafter\def\csname LTw\endcsname{\color{white}}%
      \expandafter\def\csname LTb\endcsname{\color{black}}%
      \expandafter\def\csname LTa\endcsname{\color{black}}%
      \expandafter\def\csname LT0\endcsname{\color{black}}%
      \expandafter\def\csname LT1\endcsname{\color{black}}%
      \expandafter\def\csname LT2\endcsname{\color{black}}%
      \expandafter\def\csname LT3\endcsname{\color{black}}%
      \expandafter\def\csname LT4\endcsname{\color{black}}%
      \expandafter\def\csname LT5\endcsname{\color{black}}%
      \expandafter\def\csname LT6\endcsname{\color{black}}%
      \expandafter\def\csname LT7\endcsname{\color{black}}%
      \expandafter\def\csname LT8\endcsname{\color{black}}%
    \fi
  \fi
  \setlength{\unitlength}{0.0500bp}%
  \begin{picture}(6480.00,3888.00)%
    \gplgaddtomacro\gplbacktext{%
      \csname LTb\endcsname%
      \put(990,660){\makebox(0,0)[r]{\strut{} 0.35}}%
      \put(990,1262){\makebox(0,0)[r]{\strut{} 0.36}}%
      \put(990,1863){\makebox(0,0)[r]{\strut{} 0.37}}%
      \put(990,2465){\makebox(0,0)[r]{\strut{} 0.38}}%
      \put(990,3066){\makebox(0,0)[r]{\strut{} 0.39}}%
      \put(990,3668){\makebox(0,0)[r]{\strut{} 0.4}}%
      \put(1100,440){\makebox(0,0){\strut{} 1}}%
      \put(2110,440){\makebox(0,0){\strut{} 2}}%
      \put(3120,440){\makebox(0,0){\strut{} 3}}%
      \put(4130,440){\makebox(0,0){\strut{} 4}}%
      \put(5140,440){\makebox(0,0){\strut{} 5}}%
      \put(6150,440){\makebox(0,0){\strut{} 6}}%
      \put(220,2164){\rotatebox{90}{\makebox(0,0){\strut{}$\text{BR}(B^0\to\pi^0\pi^0)[10^{-6}]$}}}%
      \put(3625,110){\makebox(0,0){\strut{}$\mu_\text{hc}[\mbox{GeV}]$}}%
    }%
    \gplgaddtomacro\gplfronttext{%
    }%
    \gplbacktext
    \put(0,0){\includegraphics{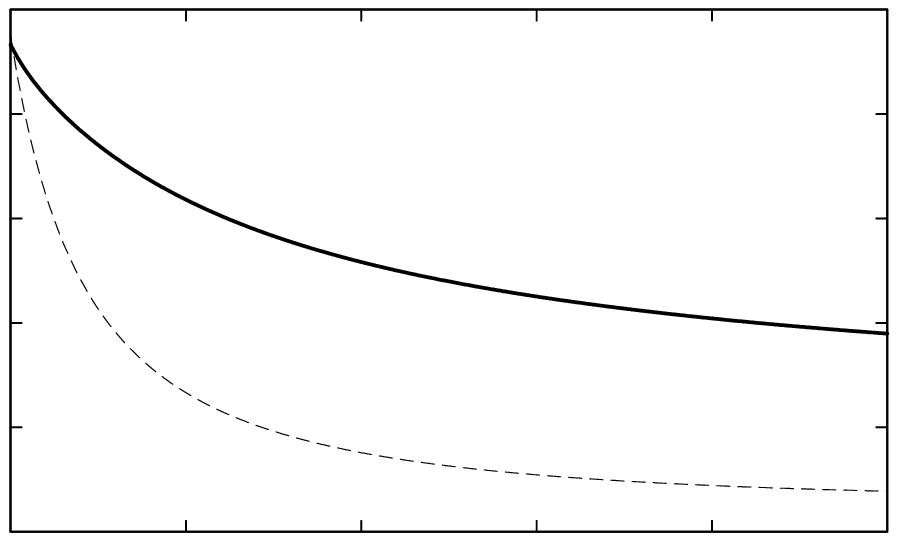}}%
    \gplfronttext
  \end{picture}%
\endgroup

%% file: conclusions/conclusions.tex
QCD factorization has turned out to be an appropriate tool to
calculate $B$ decay modes from first principles, because it allows us
to disentangle systematically the perturbative physics and the
non-perturbative physics. The present calculation showed that the hard
spectator scattering amplitude factorizes up to
$\mathcal{O}(\alpha_s^2)$, i.e.\ all infrared divergences cancel and
there are no remaining endpoint singularities. The former point is obvious
after the explicit calculation of $T^\text{II}$ and the latter point
was shown by evaluating the convolution integral (\ref{factform})
analytically. The explicit expressions for the hard spectator scattering kernel
(\ref{T1re})-(\ref{T2im}) confirmed the result of
\cite{Beneke:2005vv,Kivel:2006xc}. So they are also a confirmation that 
the leading power of the amplitudes can be obtained by performing the
power expansion at the level of Feynman integrals rather than at the
level of the QCD Lagrangian using an effective theory like SCET, which was
done in \cite{Beneke:2005vv,Kivel:2006xc}. 

The main challenges in the evaluation of Feynman integrals were due
to the fact
that the Feynman integrals came with up to five external legs and
three independent rations of scales. 
The calculation of the Feynman integrals was 
made possible with the help of tools like integration by parts
identities and differential equation techniques: 
In section \ref{diffeq} it was
shown how to get the expansion of Feynman integrals in powers of
$\lqcd/m_b$ by differential equations once the leading power is
given. 

Because next to the $m_b$-scale also the hard-collinear scale
$\sqrt{\lqcd m_b}$ enters the hard spectator scattering amplitude,
large logarithms could spoil perturbation theory. However the
numerical analysis showed a strong reduction of the scale uncertainty
of $T^\text{II}$. It also confirmed the observation of \cite{Beneke:2005vv},
that the NLO of $T^\text{II}$ is numerically important but small
enough for perturbation theory to be valid. 

Finally it is important to note that in the present calculation 
the contributions of penguin contractions and the 
effective penguin operators were not considered. 
Actually they play a dominant role in the branching ratios of 
$B\to K\pi$ and CP asymmetries of $B\to\pi\pi$ and should be taken 
into account in phenomenological applications. They have recently been  
published in \cite{Beneke:2006mk}.
Also the $\mathcal{O}(\alpha_s^2)$ corrections of $T^\text{I}$ were not part of
the present work. These contributions have been calculated in
\cite{Bell:2007tz,Bell:2007tv}.

%% file: fourpoint/fourpoint.tex
We consider the following massive four-point integral in
$d=4-2\epsilon$ dimensions (fig.~\ref{4pointfig}):
\begin{figure}
\begin{center}
\resizebox{0.5\textwidth}{!}{\includegraphics{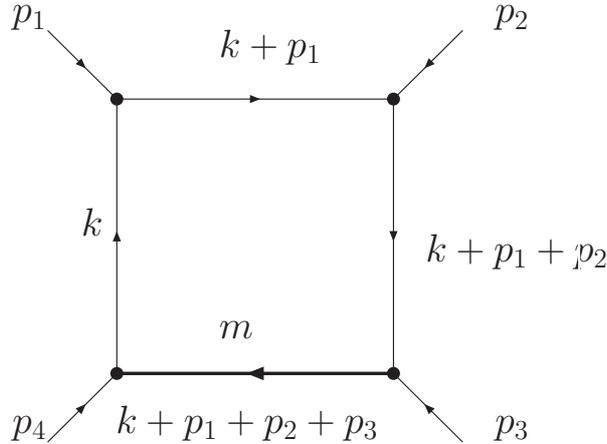}}
\end{center}
\caption{Basic one-loop four-point intergral. The massive line, which
carries the mass $m$, is indicated by the thick line.}
\label{4pointfig}
\end{figure}
\begin{equation}
I_4(p_1,p_2,p_3,p_4)=\mu^{2\epsilon}\intd
\frac{1}{D_1D_2D_3D_4}
\label{4p1}
\end{equation}
where
\begin{eqnarray}
D_1&=&k^2+i\eta
\nonumber\\
D_2&=&(k+p_1)^2+i\eta
\nonumber\\
D_3&=&(k+p_1+p_2)^2+i\eta
\nonumber\\
D_4&=&(k+p_1+p_2+p_3+p_4)^2-m^2+i\eta
\label{4p2}
\end{eqnarray}
Following \cite{Duplancic:2000sk} we introduce the external masses
\begin{equation}
p_i^2=m_i^2\quad (i=1,2,3,4)
\label{4p3}
\end{equation}
and the Mandelstam variables
\begin{equation}
s=(p_1+p_2)^2,\quad t=(p_2+p_3)^2.
\label{4p4}
\end{equation}
Furthermore we consider only the case, where 
\begin{equation}
m_2^2=0\quad \text{and}\quad m_4^2=m^2.
\label{4p5}
\end{equation}

The integral (\ref{4p1}) can be evaluated using the method of 
\cite{Duplancic:2000sk}. 
This paper gives explicit expressions for massless one-loop box
integrals. It is however possible to extend the single steps of this
paper to our case.

So finally we obtain: 
\begin{equation}
\begin{split}
&I_4(p_1,p_2,p_3,p_4)\equiv I_4(s,t,m_1^2,m_3^2,m^2)
=\frac{i}{(4\pi)^2}
\frac{\Gamma(1+\epsilon)(4\pi\mu^2)^\epsilon}{m^2(s-m_1^2)-st+m_1^2m_3^2}
\\
&\begin{split}
\quad\times\Bigg[&\frac{1}
{\epsilon}\left(\ln(-s-i\eta)+\ln(m^2-t-i\eta)
-\ln(m^2-m_3^2-i\eta)-\ln(-m_1^2-i\eta)\right)
\\
&+\ln^2(m^2-m_3^2-i\eta)+\ln^2(-m_1^2-i\eta)-\ln^2(-s-i\eta)
-\ln^2(m^2-t-i\eta)
\\
&+\ln(m^2-i\eta)\left(\ln(-s-i\eta)+\ln(m^2-t-i\eta)
-\ln(m^2-m_3^2-i\eta)-\ln(-m_1^2-i\eta)\right)
\\
&+2\li\left(1-\frac{m^2-t-i\eta}{-m_1^2-i\eta}\right)
-2\li\left(1-\frac{m^2-m_3^2-i\eta}{-s-i\eta}\right)
\\
&+2\li\left(1-(m_3^2-m^2+i\eta)f^m\right)
+2\li\left(1-(m_1^2+i\eta)f^m\right)
\\
&-2\li\left(1-(t-m^2+i\eta)f^m\right)
-2\li\left(1-(s+i\eta)f^m\right)
\Bigg],
\end{split}
\end{split}
\label{4p6}
\end{equation}
where
$f^m=\frac{s+t-m_1^2-m_3^2}{m^2(m_1^2-s)+st-m_1^2 m_3^2}$.

The case $m_1^2=0$ gives rise to further divergences and has to be
considered separately:
\begin{equation}
\begin{split}
&I_4(s,t,m_1^2=0,m_3^2,m^2)
=\frac{i}{(4\pi)^2}
\frac{\Gamma(1+\epsilon)(4\pi\mu^2)^\epsilon}{s(m^2-t)}
\\
&\begin{split}
\quad\times\Bigg[&
-\frac{3}{2\epsilon^2}+
\frac{1}
{\epsilon}\left(2\ln(m^2-t-i\eta)-\frac{1}{2}\ln(m^2-i\eta)
+\ln(-s-i\eta)-\ln(m^2-m_3^2-i\eta)\right)
\\
&+\frac{2\pi^2}{3}
+\frac{1}{4}\ln^2(m^2-i\eta)-\ln^2(m^2-t-i\eta)+\ln^2(m^2-m_3^2-i\eta)
-\ln^2(-s-i\eta)
\\
&+\ln(m^2-i\eta)\left(\ln(-s-i\eta)-\ln(m^2-m_3^2-i\eta)\right)
\\
&-2\li\left(1-\frac{m^2-m_3^2-i\eta}{-s-i\eta}\right)
+2\li\left(1-(m_3^2-m^2+i\eta)f^m\right)
\\
&-2\li\left(1-(t-m^2+i\eta)f^m\right)
-2\li\left(1-(s+i\eta)f^m\right)
\Bigg],
\end{split}
\end{split}
\label{4p7}
\end{equation}
where
$f^m=\frac{s+t-m_3^2}{s(t-m^2)}$.